\documentclass{emulateapj}
\shorttitle{Cluster formation and anisotropic thermal conduction}
\shortauthors{Ruszkowski et al.}

\newcommand{\brunt}{Brunt-V\"ais\"al\"a \,}
\def\gsim{\;\rlap{\lower 2.5pt
 \hbox{$\sim$}}\raise 1.5pt\hbox{$>$}\;}
\def\lsim{\;\rlap{\lower 2.5pt
   \hbox{$\sim$}}\raise 1.5pt\hbox{$<$}\;}

\begin{document}

\title{Cosmological MHD simulations of cluster formation with anisotropic \\
thermal conduction}

\author{M. Ruszkowski$^{1,2}$, D. Lee$^{3}$, M. Br{\"u}ggen$^{4}$, I. Parrish$^{5}$, and S.Peng Oh$^{6}$}
\affil{$^{1}$Department of Astronomy, University of Michigan, 500 Church Street, Ann Arbor, MI 48109, USA; 
e-mail: mateuszr@umich.edu}
\affil{$^{2}$The Michigan Center for Theoretical Physics, 3444 Randall Lab, 450 Church St, Ann Arbor, MI 48109, USA}
\affil{$^{3}$Department of Astronomy, ASC/Flash Center, University of Chicago, 5640 S. Ellis Avenue, Chicago, IL 60637, USA; 
e-mail: dongwook@flash.uchicago.edu}
\affil{$^{4}$Jacobs University, Bremen, Campus Ring 1, 28759 Bremen 05233, Germany; e-mail: m.brueggen@jacobs-university.de}
\affil{$^{5}$Astronomy Department and Theoretical Astrophysics Center, 601 Campbell Hall, University of California, Berkeley, CA 94720, USA; e-mail: iparrish@astro.berkeley.edu}
\affil{$^{6}$Department of Physics, University of California, Santa Barbara, CA 93106, USA; e-mail: peng@physics.ucsb.edu}

\begin{abstract}

The intracluster medium (ICM) has been suggested to be 
buoyantly unstable in the presence of magnetic field and anisotropic thermal conduction.
We perform first cosmological simulations of galaxy cluster formation that simultaneously include magnetic fields, radiative cooling 
and anisotropic thermal conduction. In isolated and idealized 
cluster models, the magnetothermal instability (MTI) tends to reorient the magnetic fields radially whenever the temperature 
gradient points in the direction opposite to gravitational acceleration.
Using cosmological simulations of the {\it Santa Barbara} cluster we detect radial bias in the velocity and magnetic fields. 
Such radial bias is consistent with either the inhomogeneous radial gas flows due to substructures or residual MTI-driven field rearangements
that are expected even in the presence of turbulence. Although disentangling the two scenarios is challenging, we do not
detect excess bias in the runs that include anisotropic thermal conduction.
The anisotropy effect is potentially detectable via radio polarization measurements with {\it LOFAR} and {\it Square Kilometer Array} 
and future X-ray spectroscopic studies with the {\it International X-ray Observatory}. 
We demonstrate that radiative cooling boosts the amplification of the magnetic field by about two orders of magnitude
beyond what is expected in the non-radiative cases. This effect is caused by the compression of the gas and frozen-in magnetic field 
as it accumulates in the cluster center. At $z=0$ the field is amplified by a factor of about $10^{6}$ compared to 
the uniform magnetic field evolved due to the universal expansion alone. 
Interestingly, the runs that include both radiative cooling and anisotropic thermal conduction exhibit stronger magnetic 
field amplification than purely radiative runs, especially at the off-center locations.
In these runs, shallow temperature gradients away from the cluster center 
make the ICM neutrally buoyant. Thus, the ICM is more easily mixed in these regions and the winding up 
of the frozen-in magnetic field is more efficient resulting in stronger magnetic field amplification.
We also demonstrate that thermal conduction partially reduces the gas accretion driven by overcooling
despite the fact that the effective conductivity is suppressed below the Spitzer-Braginskii value.

\end{abstract}


\section{Introduction}

Thermal conduction may play an important role in the evolution of the intracluster medium (e.g., \citet{voigt04}). 
Sufficiently strong conduction can offset radiative cooling losses in massive clusters and reduce the energy requirements
on active galactic nuclei (AGN) feedback that is required to prevent overcooling in less massive clusters and groups
(see \citet{mcnamara07} and \citet{norman} for reviews).
It has also been suggested that not only can thermal conduction serve as a mechanism for cool core heating but that it is very important 
for the stability of these systems (\citet{ruszkowski02}, \citet{guo08b}, Ruszkowski \& Oh, in prep.). 
It may also be responsible 
for setting a critical central entropy threshold below which star formation is possible in cluster cool cores \citep{voit08}.
The recently discovered bimodality in the distribution of the cluster central entropy 
\citep{cavagnolo09, sanderson09}
may be due to the combination of AGN feedback from the brightest cluster galaxies that stabilizes low entropy clusters and 
a combination of mergers and thermal conduction that 
stabilize higher central entropy clusters \citep{guo08b, ruszkowski10, parrish10}.
As such, thermal conduction may be important for understanding of the feeding of the most massive black holes in the 
Universe. In cluster outskirts thermal conduction may flatten
the temperature distributions \citep{lemaster}, which may have consequences for the cluster mass estimates. This may have
possible impact for precision cosmology as it relies on accurate mass measurements in  the most massive clusters.\\
\indent
As a plasma transport process, thermal conduction is closely linked to gas viscosity. Both types of 
transport processes may explain various X-ray observations. For example, recent {\it Chandra} observations of 
M87 \citep{werner10} show that the temperature in the shells centered on the cluster center is remarkably isothermal. They suggest that 
such a high degree of isothermality is consistent with 
effective heat conduction in the tangential direction. Moreover, they also 
attribute
the presence of small scale metallicity gradients to relatively weak level of small scale turbulence, which could be 
consistent with 
viscous damping of gas motions. AGN are known to generate intermittent outflows causing weak shocks and sound waves
(e.g., \citet{fabian_shocks},  \citet{forman}, \citet{finoguenov}). The dissipation of the 
energy contained in these waves may be sufficient to offset radiative cooling of the gas. It has been shown observationally 
(in {\it Perseus}: \citet{fabian_shocks}; in {\it Virgo}: \citet{forman}) and using numerical simulations 
\citep{ruszkowski04a, ruszkowski04b, bruggen05}
that Spitzer-Braginskii viscosity and/or conduction is sufficient to dissipate such waves and heat the ICM efficiently. Moreover, the compact 
morphology of the buoyantly rising AGN is consistent with the presence of effective viscosity \citep{reynolds}
although it could also be explained by the magnetic draping effect \citep{robinson,ruszkowski07,ruszkowski08,dursi,oneill09}. 
Otherwise, the bubbles would be disrupted by Rayleigh-Taylor instabilities. Gas viscosity has also been shown to play a role in shaping 
the properties of cold fronts caused by the sloshing motion (e.g., \citet{zuhone}).\\
\indent
The simple picture of viscosity and conduction is complicated by the presence of magnetic fields that are known to be present 
in the ICM \citep{ensslin03,vogt03,feretti}.
Magnetic fields suppress thermal conduction in the direction perpendicular to the B-field. However, even in the case of highly 
tangled magnetic fields, the effective thermal conduction can be a substantial fraction the Spitzer conductivity \citep{narayan01}.
The anisotropy due to magnetic fields leads to new phenomena with potentially important consequences for the 
ICM. It has been demonstrated both analytically \citep{balbus00} and numerically \citep{parrish05}
that the ICM is unstable in the presence of a weak magnetic field 
and anisotropic thermal conduction when the temperature is increasing in the direction of gravity
(magnetothermal instability, MTI). When magnetic fields are
partially aligned with the temperature gradient, the heat flow from hotter to cooler regions 
along the field lines makes such regions more buoyant. This causes the magnetic field to become 
more aligned with the temperature gradient and leads to 
the instability and preferentially radial magnetic fields. 
A similar instability occurs when anisotropic transport of energy via cosmic rays takes place \citep{chandran06}. 
More recent analysis shows that the gas is also unstable
when the temperature decreases in the direction of gravity
in the presence of a background heat flux (HBI instability; \citet{quataert08}) which has also been verified 
by numerical simulations \citep{parrish08a}. The saturated state of the HBI corresponds to the magnetic fields 
oriented in the direction perpendicular to gravity. Such field configuration implies effectively vanishing thermal conduction
from the hotter outer cluster layers to their cool cores. This accelerates the effective cooling rate in the core. 
However, the exact topology of the magnetic fields depends also on whether externally imposed turbulence driving is present.
Such turbulence may come from AGN outbursts, galaxy motions and structure formation (major and minor mergers). This 
has been recently investigated by \citet{ruszkowski10} and \citet{parrish10} who showed that there exists a critical 
level of turbulence above which the field can be randomized and the conductive heating to the core restored. Although this analysis
was performed for the HBI, similar arguments apply to the MTI.\\
\indent
The exact level of turbulence in the ICM is not known but is expected to vary throughout the evolution of a cluster. However, 
indirect measurements (e.g., \citet{churazov_velocity}) and upper limits (\citet{sanders_velocity}) are available.
For example,
some observational estimates put the ICM velocity at a level of local sound speed \citep{markevitch02, mahdavi_velocity},  
while others suggests relatively ``calm'' ICM \citep{werner10}. These levels of turbulence are expected theoretically 
(e.g., \citet{evrard}, \citet{nagai03}, \citet{vazza}). Future measurements with the {\it International X-ray Observatory} 
(IXO) will help to determine the level of turbulence in clusters more precisely \citep{heinz_xim}.
Therefore, ideally, we need to resort to ab initio cosmological simulations 
to include the effects of structure formation. Using this approach we can not only
compute the level of the effective thermal conduction in the presence of magnetic fields but we can also simulate the growth of magnetic 
field. Both MTI and HBI are expected to amplify the fields due to kinematic dynamo action although the efficiency of this process is modest. 
However, trapping of gravity modes
may lead to vorticity growth \citep{lufkin95} and further field amplification \citep{ruszkowski10b}.\\
\indent
Here we present first cosmological simulations of cluster formation that simultaneously include radiative cooling, magnetic field and 
anisotropic thermal conduction. These simulations are a natural extension of our previous work on the role of 
radiative cooling, magnetic fields, conduction and viscosity with the {\it FLASH} and 
{\it ATHENA} codes that focused on isolated cool cores \citep{ruszkowski04a, ruszkowski04b, bruggen05, bogdanovic09, ruszkowski10, ruszkowski10b, parrish10}.  
This work also builds on previous theoretical and numerical efforts to 
simulate the growth of magnetic field in cluster formation simulations by our group
\citep{bruggen05b}, with the {\it FLASH} code and other teams \citep{dolag99, dolag02}
with {\it GADGET}; \citep{collins} {\it Enzo}, 
\citep{li} {\it CosmoMHD}, \citep{dubois}, {\it RAMSES} as well as the work by \citet{sijacki}, 
who considered non-MHD simulations but included viscosity, and \citet{jubelgas04} and \citet{dolag04}
who considered conduction in hydrodynamical SPH simulations.
We also note that the field amplification may result from purely kinetic plasma processes (e.g., \citet{scheko07}, \citet{scheko10}, \citet{kunz})
and that the large scale turbulence may serve as the energy input for these mechanisms.
However, these processes are beyond the scope of this investigation and are not considered here.\\
\indent
The paper is organized as follows. In the next section we discuss the methods and the code used for the simulations.
In Section 3 we discuss our results. Section 4 presents the conclusions. Appendix A discusses the comparison
of the anisotropic conduction module with the linear theory predictions. In Appendix B we briefly discuss the tests of the implementation of the
cosmological terms in the MHD equations.

\section{Methods}

We solve the following set of MHD equations augmented by the cosmological expansion terms.

\begin{eqnarray}
\frac{\partial\rho}{\partial t}+\nabla\cdot(\rho{\bf v}) = 0 
\end{eqnarray}
\begin{eqnarray}
\frac{\partial(\rho{\bf v})}{\partial t}+\nabla\cdot(\rho{\bf vv-BB})+\nabla p = -\rho{\nabla\phi} -2\frac{\dot{a}}{a}\rho{\bf v} 
\end{eqnarray}
\begin{eqnarray}
\frac{\partial E}{\partial t}+\nabla\cdot({\bf v}(E+p)-{\bf B}({\bf v}\cdot{\bf B})) = -\rho\nabla\phi\cdot{\bf v} \nonumber \\ 
+ \frac{\dot{a}}{a} \left[ (3\gamma -1)\rho\epsilon +2\rho v^{2}\right] + a{\cal C}(T_{\rm ph}) - \nabla\cdot{\bf F}_{\rm ph}
\end{eqnarray}
\begin{eqnarray}
\frac{\partial{\bf B}}{\partial t} + \nabla\cdot ({\bf vB-Bv}) = 0, 
\end{eqnarray}

\noindent
where
\begin{eqnarray}
p=p_{\rm th}+\frac{B^{2}}{2}
\end{eqnarray}
\begin{eqnarray}
E=\frac{\rho v^{2}}{2}+\epsilon +\frac{B^{2}}{2}
\end{eqnarray}
\begin{eqnarray}
\Delta\phi = 4\pi Ga^{-3}(\rho -\langle\rho\rangle),\\ \nonumber
\end{eqnarray}

\noindent
where $\langle\rho\rangle$ is the comoving mean density and
where $p_{\rm th}$ is the gas pressure and $\epsilon$ is the gas internal energy per unit volume. We assume the
adiabatic index $\gamma = 5/3$ and the mean molecular weight $\mu = 0.5$ in the equation of state. 
In equation (3), ${\cal C}$ represents the cooling rate per unit volume. 
We use standard tabulated and publicly available cooling curves \citep{sutherland93} for metallicity $Z = 0.3Z_{\odot}$.
Our simulations do not include star formation or AGN feedback and we set a floor in the cooling function at 0.01 keV in physical units.\\
\indent
The anisotropic thermal conduction heat flux ${\bf F}_{\rm ph}$ is given by

\begin{eqnarray}
{\bf F}_{\rm ph}=-a^{-1}\kappa (T_{\rm ph}) \hat{\bf e}_{B}(\hat{\bf e}_{B}\cdot{\mathbf\nabla} T_{\rm ph}),\\ \nonumber
\end{eqnarray}

\noindent
where $\hat{\bf e}_{B}$ is a unit vector pointing in the direction of the magnetic field and $\kappa$ is the Spitzer-Braginskii 
conduction coefficient given by $\kappa (T) = 4.6\times 10^{-7}T^{5/2}$erg s$^{-1}$cm$^{-1}$K$^{-1}$.
In the above equation and in all equations below all variables with the ``ph'' subscript denote physical quantities.
Following \citet{cowie77} 
we included the effect of conduction saturation whenever the characteristic lengthscale associated with the temperature gradient 
exceeds the mean free path, though in the bulk of the ICM this effect is not significant.
We also imposed an upper limit on conduction such that $K\equiv a^{-6}\kappa /(c_{v}\rho)$, where
$c_{v}^{-1} = (\gamma -1)\mu m_{\rm prot}/k_{\rm boltz}$ (the extra factor of $a$ comes from the prefactor in Equation 8). 
We set $K<K_{\max} = 5\times 10^{32}$cm$^{2}$s$^{-1}$.
This ceiling was introduced in order to prevent extremely small diffusive timesteps.
We found that changing $K_{\rm max}$ to larger values did not affect our results significantly. Only a relatively limited volume
far from the cluster center was subject to this upper limit.\\
\indent 
We also note that the effective conduction may be limited in the outer regions of the cluster for physical reasons \citep{medvedev07}. 
Moreover, around or beyond the virial radius, the electron-proton energy equilibration timescales start to exceed the dynamical or buoyancy 
timescale. In this region, the MHD approximation starts to break down and the MTI growth rates are reduced. 
However, at smaller radii, the equilibration timescale due to Coulomb collisions are  
comparable to or shorter than the buoyancy timescale, 
some thermal coupling between electrons and protons begins to be possible, and the MHD analysis applies
at least approximately. Finally, there are independent observational arguments for fast equilibration in the ICM. Specifically, 
\citet{markevitch07} detected a sharp electron temperature jump in the post shock region in the Bullet Cluster 
and used this observation to suggests that the electron-proton equilibration is much shorter than the collisional timescale.
The inferred equilibration timescale in the postshock ICM 
is at least 5 times shorter than the Coulomb equilibration time and may even be consistent with being instantaneous.
While the issue of equilibration is clearly
an open research topic, here we work under the symplifying assumption that equilibration is instantaneous and
that the MHD approximation is appropriate, and then investigate the consequences of these assumptions.\\
\indent
The most stringent limitation 
on the timestep $dt$ in the simulation was due to the diffusive process with $dt\sim (\Delta x)^{2}/K$. 
The maximum resolution that we could afford computationally was 7 levels of refinement for blocks consisting of 16 zones on a side.
That is, the effective resolution was $\sim 31h^{-1}$kpc. The strong limit imposed on the timestep can be avoided by employing 
implicit integration methods. We are now implementing such methods in the {\it FLASH} code (Lee et al., in prep.).
This approach will significantly speed up the computations and will allow us to perform simulations for a range of cluster masses
(Ruszkowski et al., in prep).\\
\indent
Equations 1 through 8 were obtained by starting from the MHD equations and applying cosmological expansion transformation for the
spatial gradients and the following variable transformations

\begin{eqnarray}
\rho = a^{3}\rho_{\rm ph}
\end{eqnarray}
\begin{eqnarray}
p = ap_{\rm ph}
\end{eqnarray}
\begin{eqnarray}
T=a^{-2}T_{\rm ph}
\end{eqnarray}
\begin{eqnarray}
\rho\epsilon = a\rho_{\rm ph}\epsilon_{\rm ph}
\end{eqnarray}
\begin{eqnarray}
{\bf B} = a^{1/2}{\bf B}_{\rm ph},\\ \nonumber
\end{eqnarray}

\noindent
where $a$ is the cosmological expansion factor.
The velocity vector ${\bf v}=\dot{{\bf x}}$, where ${\bf x}=a^{-1}{\bf r}$ is the code position vector 
and ${\bf r}$ is the physical position vector.
These transformations together with Equations 1 to 7 lead to correct scalings of all quantities with the expansion factor. 
For example, $T_{\rm ph}\propto a^{-2}$ and $B_{\rm ph}\propto a^{-2}$ which ensures the conservation of the magnetic flux.
Note that the variable transformation adopted here is somewhat different than, e.g., that in \citet{li} in the {\it CosmoMHD} code,
but the scalings of the physical quantities with redshift is correct in both cases.
The cosmological terms were included using the operator splitting technique. Specifically, we 
computed the updates to all variables due to the cosmological expansion terms by finding the exact solutions 
to a modified set of Equations 1 to 4 that retained only the time derivative terms.
Tests of the implementation of the cosmological terms are presented in Appendix A.\\
\indent
We used publicly available initial conditions for the Santa Barbara cluster \citep{frenk99}. 
These conditions correspond to initial 3$\sigma$ density perturbations spread over 10 Mpc, Hubble constant 
of 50 km/s/Mpc and flat geometry in a matter dominated universe ($\Lambda = 0$) and result in the formation of 
a cluster characterized by $r_{\rm vir}\sim 2.7$Mpc. This choice of publicly available 
initial conditions allows for standardized comparisons with other codes. We employed periodic boundary conditions.
We note that the initial conditions are not sensitive to the initial temperature and strength of the magnetic fields as long 
as they are small. The initial magnetic field was assumed to be constant in space. 
While in isolated cluster simulations the initial topology of the magnetic field can affect 
the instability growth rate, the frozen-in magnetic field in the cosmological simulations is
randomized by structure formation motions well before the hot virialized and relaxed structures 
are formed. Since early on thermal conduction is not expected to be very important, 
the initial field topology is not expected to play a crucial role. In fact, it has been demonstrated using cosmological MHD simulations 
without transport processes that  
the results were indeed not sensitive to the exact topology of the magnetic field in the statistical sense
\citep{dolag02, bruggen05b}. Therefore, for simplicity we assume initial magnetic field that is pointing 
in the same direction and is uniform. We set the initial strength of the 
magnetic field to $10^{-11}/(4\pi)^{1/2}$ Gauss, i.e., the initial physical magnetic field strength 
was $10^{-11}/(4\pi)(1+z)^{1/2}$ Gauss at the initial redshift of $z=20$.\\
\indent
The simulations were performed with the adaptive mesh refinement (AMR) {\it FLASH} code.
{\it FLASH} is a publicly available code that 
was in part developed by the DOE-supported ASC/Alliance Center for Astrophysical Thermonuclear Flashes at the University of Chicago.
It is a modular, parallel simulation code capable of handling general compressible flow
problems found in many astrophysical environments. The code is parallelized using
Message-Passing Interface (MPI) library and the HDF5 
or PnetCDF library for parallel I/O to achieve portability and scalability on a variety of different parallel computers.\\
\indent
The new directionally unsplit staggered mesh MHD solver (USM; \citet{lee09a,lee09b})
is based on a finite-volume, high-order Godunov method combined with a constrained transport (CT)
scheme which ensures divergence-free magnetic fields. Tests of the module demonstrate
that it is very robust and significantly outperforms the previously implemented MHD 8-wave
solver in {\it FLASH}.\\
\indent
We implemented the anisotropic conduction unit
following the approach of \citet{SharmaHammett07}. More specifically, we applied monotonized central (MC) limiter to the conductive fluxes.
This method ensures that anisotropic conduction does not lead to negative temperatures in the presence of steep temperature 
gradients. We verified that the module predicts correct magnetothermal instability growth rates. Details of these tests are 
discussed in Appendix B.\\

\section{Results}

\begin{figure*}
\begin{center}
\includegraphics[width=0.9\textwidth]{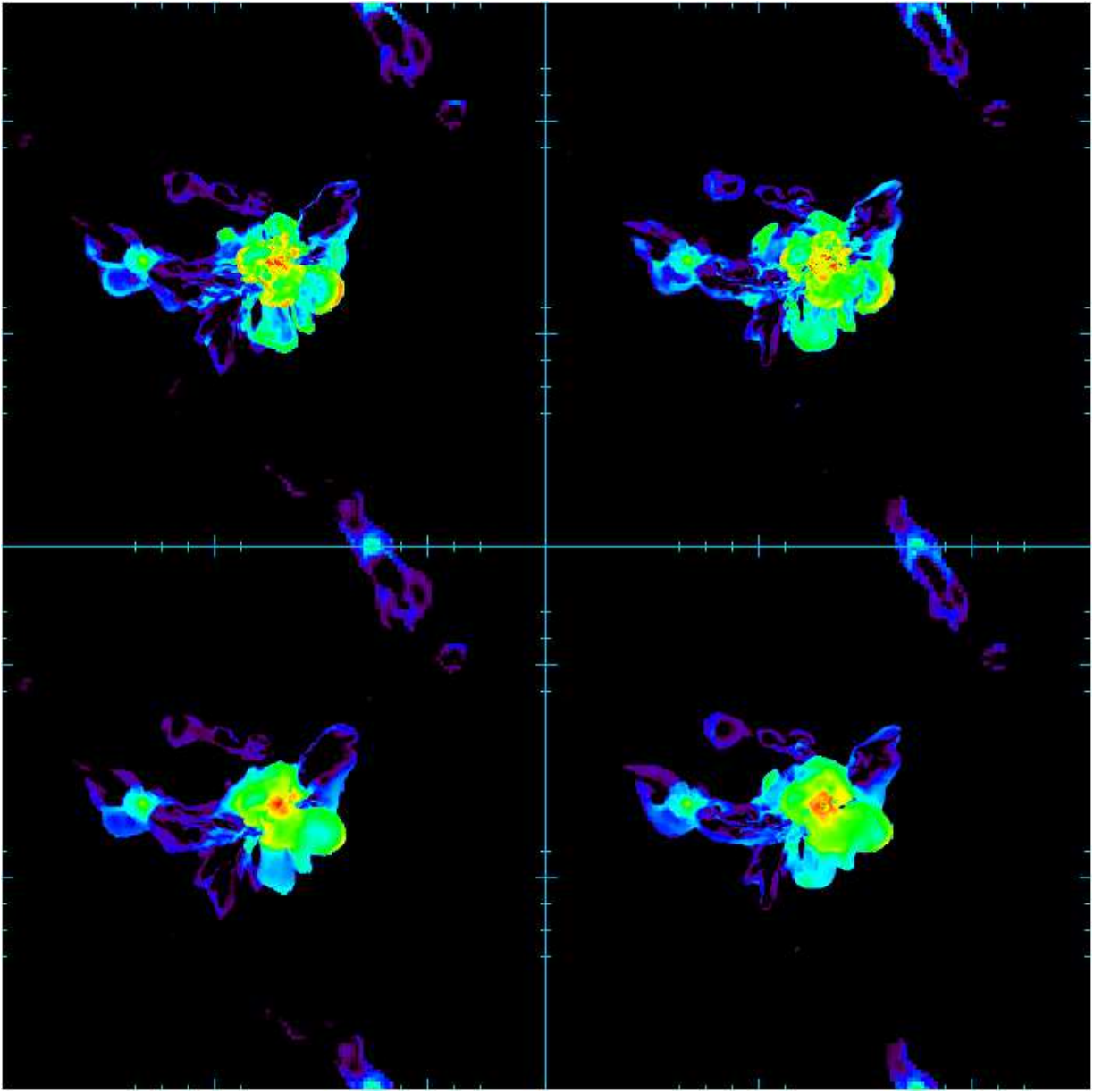}
\end{center}
\caption{Cross sections through the temperature distribution in the cluster at $z=0$. All panels correspond to $32h^{-1}$ Mpc
on a side. The top row is for the non-conductive cases while the bottom row is for the runs with anisotropic thermal conduction.
Left columns corresponds to non-radiative simulations and the right one is for the runs with radiative cooling.
The minimum and maximum values of the temperature are the same in all panels. 
This figure illustrates the relative differences between simulations including different physics processes. 
Their effect is quantified in Figure 2 that shows temperature profiles.\\}
\end{figure*}

\subsection{Temperature distribution}

Figure 1 shows cross sections through the temperature distribution in the cluster. All panels correspond to $32h^{-1}$Mpc
on a side. The minimum and maximum values of the temperature are the same in all panels. 
Top row is for the non-conductive cases while the bottom row is for the runs with anisotropic thermal conduction.
Left columns is for non-radiative simulations and the right one for the runs with radiative cooling. The upper right panel 
clearly shows a well-developed cool core. Comparison of the top and bottom rows shows that conduction is efficient in smearing out the
fine structure details in the temperature distribution. The cool core in the conductive case is partially heated by 
thermal conduction from hotter outer layers of the cluster but it does not become isothermal. 

\begin{figure}
\begin{center}
\includegraphics*[viewport = 70 0 500 500, width=0.5\textwidth]{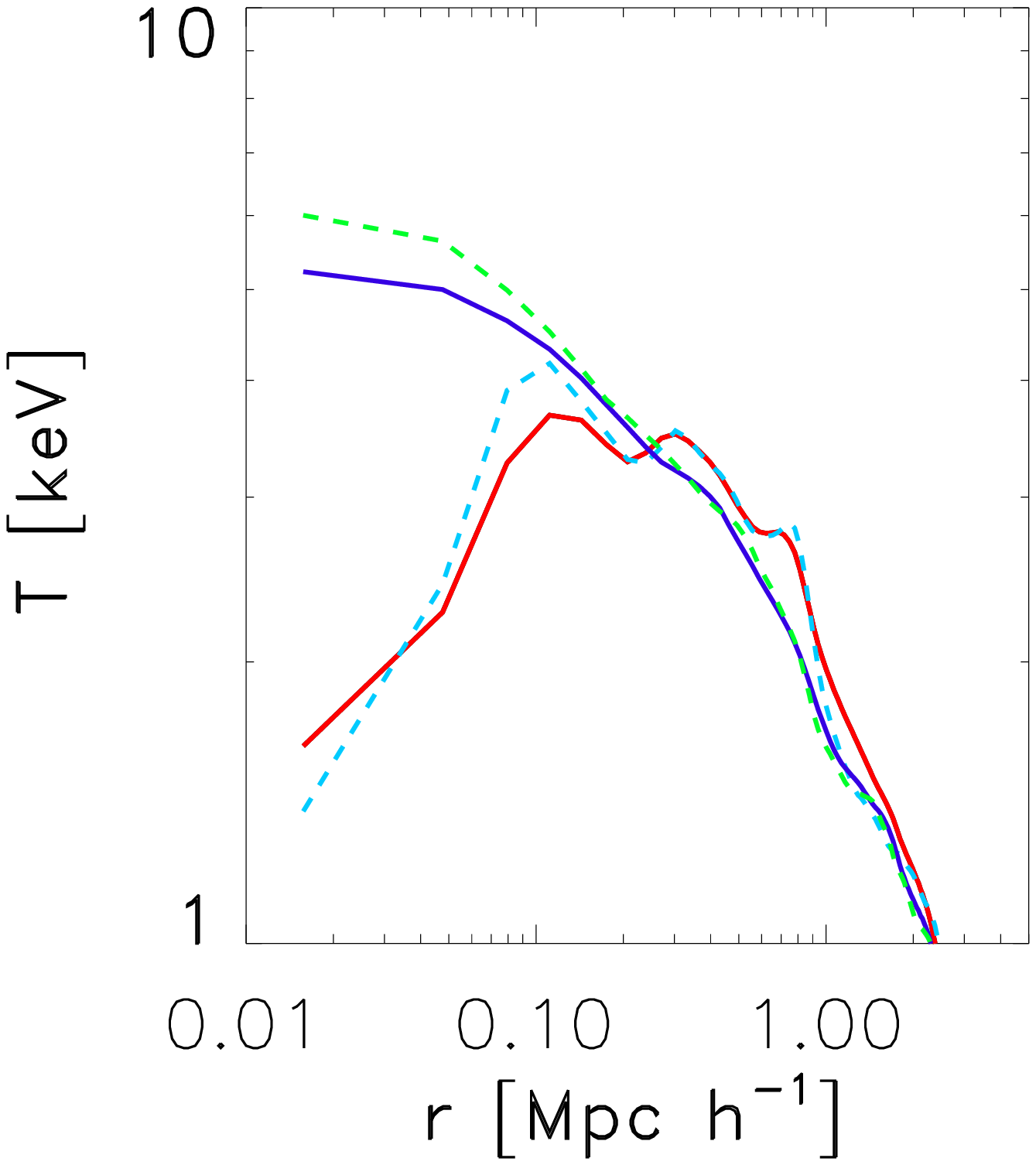}
\end{center}
\caption{Temperature profiles of the cluster. Dashed green curve is for the
adiabatic run, solid dark blue for anisotropic conduction, dashed light blue for radiative cooling, and 
solid red is for radiative cooling with anisotropic conduction. The profiles are not density-weighted.\\}
\end{figure}

\indent
In Figure 2 we show the temperature profiles of the cluster for all four cases shown in Figure 1. 
The profiles are not density-weighted. Dashed green curve is for the
non-radiative run, solid dark blue for anisotropic conduction, dashed light blue for radiative cooling, and 
solid red is for radiative cooling with anisotropic conduction. The adopted bin size was 
$\sim 31h^{-1}$ kpc.
For larger bin sizes, the comparison between the temperature profiles and 
the temperature maps shown in Figure 1 is more difficult; other one-dimensional profiles presented below use a bin size of 
$125h^{-1}$ kpc, 
which is especially important for the observables inferred from the vector quantities as they tend to exhibit larger fluctuations.
The effect of thermal conduction on the temperature profiles is mild but noticeable. 
This is partially due to the fact that magnetic fields reduce the effective level of 
conduction below the Spitzer value. The effect is more pronounced in \citet{jubelgas04} who use full isotropic 
Spitzer conductivity in cosmological simulations.
Radiative cooling has much stronger effect on the temperature distribution. Interestingly,
the cooling runs show excess temperature at larger radii. We interpret this effect as a consequence of 
the combination of the increased role of shocks in the cooling ICM that generate entropy and 
``adiabatic'' compression (e.g., \citet{lufkin00}). Both of these effects are more important in the presence of radiative losses in the 
cluster center. The rapid cooling ``pulls'' the 
outer cluster layers toward the center and heats up the ICM. In the presence of cooling the shocks are stronger and the heating of the
gas more efficient. This effect is illustrated in Figure 3 where we show the entropy profiles. The meaning of the curves is the same
as in Figure 2. Figure 3 shows that the radiative runs exhibit elevated entropy at $r\ga 300h^{-1}$kpc.

\begin{figure}
\begin{center}
\includegraphics*[viewport = 70 0 500 500, width=0.5\textwidth]{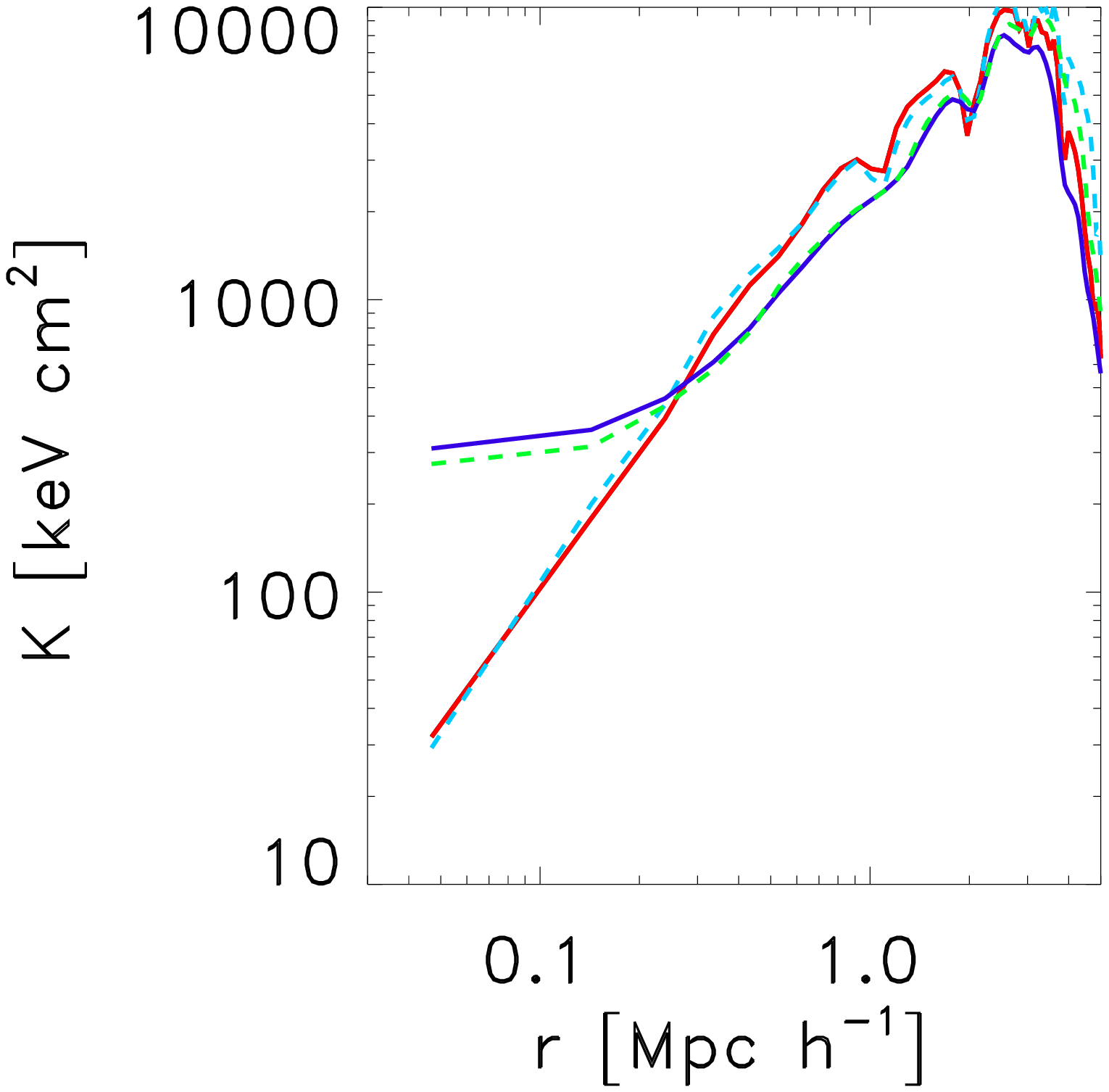}
\end{center}
\caption{Entropy profiles of the cluster.  The meaning of the curves is the same as in Figure 2. Entropy is defined as 
$K=k_{\rm boltz}T/n_{e}^{2/3} $ and the plot units are [keV cm$^2$].\\}
\end{figure}

We point out that the temperature profile declines with the distance from the cluster center even in the runs that
include conduction. This is possible when the virialization shocks heat the gas faster than conduction can remove the heat.

\begin{figure*}
\includegraphics[width=0.33\textwidth]{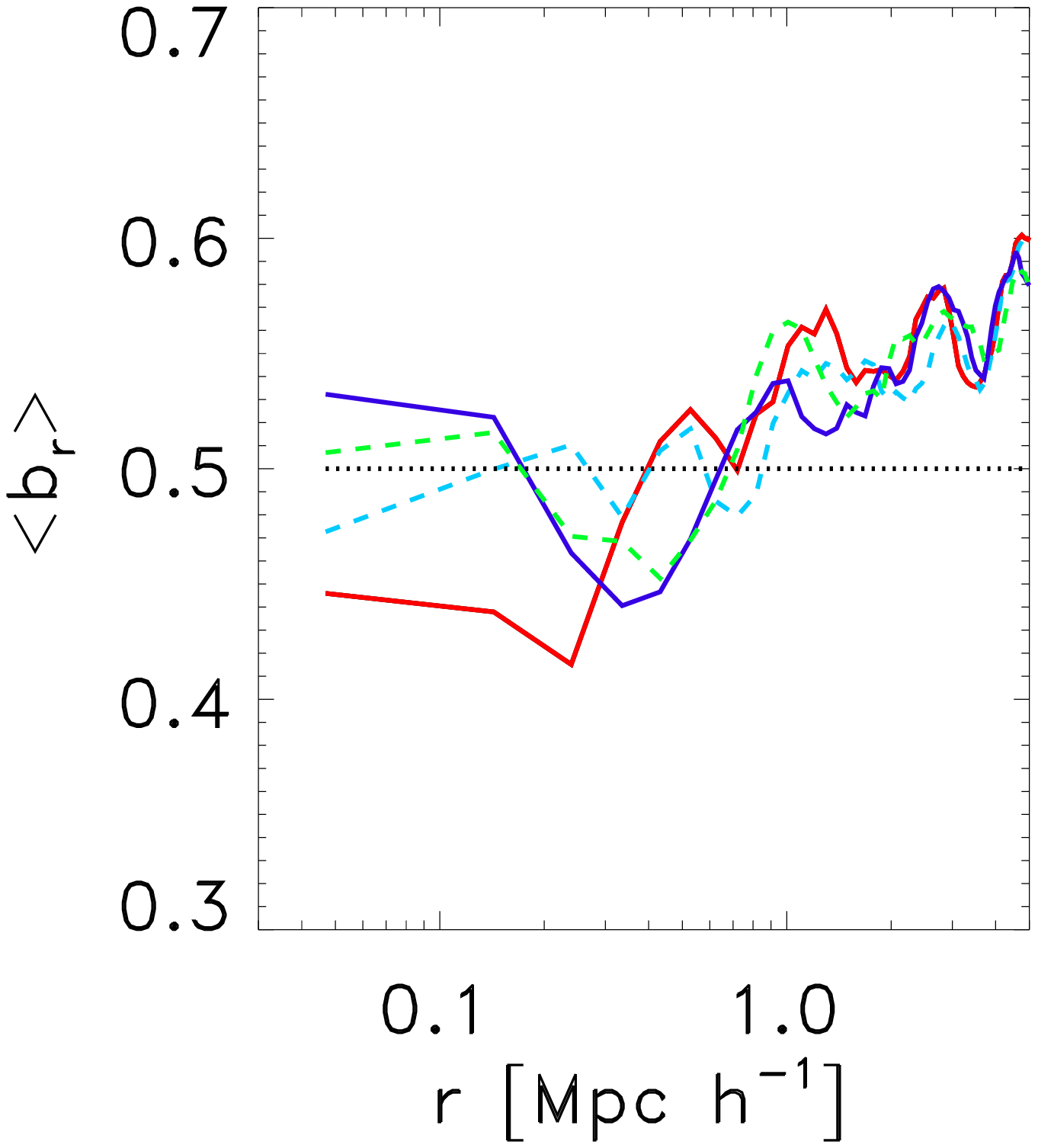}
\includegraphics[width=0.33\textwidth]{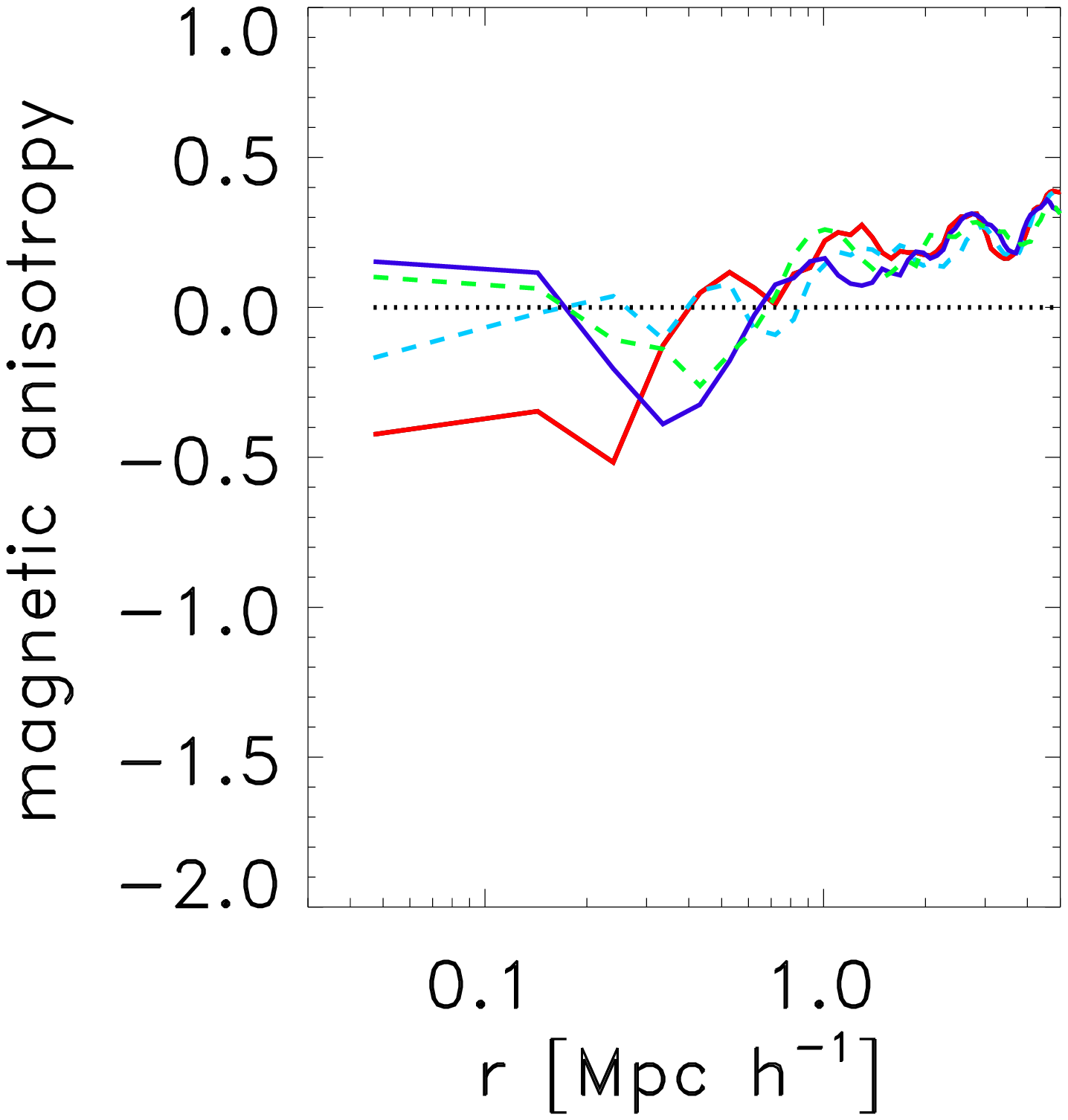}
\includegraphics[width=0.33\textwidth]{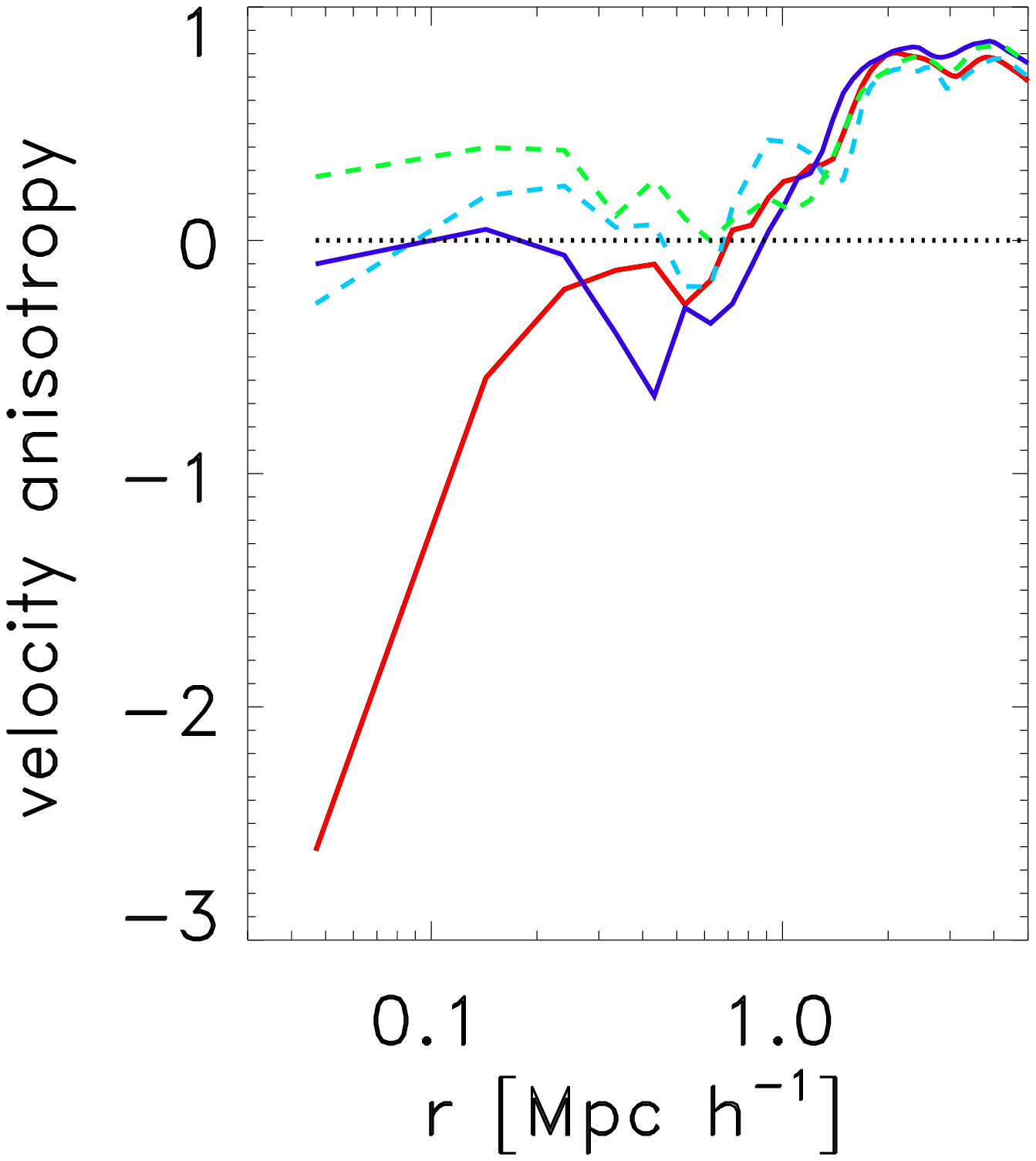}
\caption{Left panel: absolute value of the radial component of the magnetic field vector. Middle and right panels show anisotropy parameters
for the magnetic (middle panel) and velocity fields. Negative values correspond to tangential fields, zero is for 
the isotropic case. Anisotropy is defined as $\beta = 1-\sigma^{2}_{t}/2\sigma^{2}_{r}$, where $\sigma_{r}$ and $\sigma_{t}$
are the dispersions of either the magnetic field or the velocity field. The color coding of all curves is the same as in Figure 2.
Horizontal dotted lines correspond to isotropic configurations.\\}
\end{figure*}

\subsection{Statistical properties of magnetic and velocity field orientations}

\indent
In the very central parts of the cluster the suppression of the effective conduction is further enhanced by 
partially tangential ordering of the fields. This effect is shown in the left panel in 
Figure 4, where we plot the absolute value of the radial component of the magnetic field vector (left panel) and
anisotropy parameter $\beta$
for the magnetic field (middle panel). This quantity is defined as $\beta = 1-\sigma^{2}_{bt}/2\sigma^{2}_{br}$, where
$\sigma_{bt}$ and $\sigma_{br}$ are the transverse and radial magnetic field dispersions, respectively. 
For example, $\sigma_{bt}=\langle B_{t}^{2}\rangle^{1/2}$ and the definition of $\sigma_{br}$ is analogous. The meaning of 
the curves is the same as in Figure 2. 
For the absolute value of the radial component of the magnetic field, the isotropic case corresponds to 0.5. For magnetic anisotropy,
isotropic case corresponds to vanishing $\beta$, and tangential and radial cases to negative and positive $\beta$, respectively.
Horizontal dotted lines correspond to isotropic configurations.
Figure 4 demonstrates that the radiative run with field-aligned thermal conduction 
shows a weak bias for tangential orientation of the magnetic fields. 
We stress that this feature, although technically corresponding to a region resolved by up to 16 grid zones, is
only tentative and that higher resolution runs are required to make any definite statements about its nature.
In a companion paper (Ruszkowski \& Oh 2010b, submitted)
we make a step in this direction by systematically studying the substructure parameter space and the substructure impact on 
the HBI in isolated cool cores in non-cosmological simulations.
There is also a tentative decrement in the anisotropy parameter at intermediate radii for all four runs. 
This may be caused by partial trapping of gravity modes \citep{rebusco08, ruszkowski10}. 
At larger radii there appears to be a 
slight radial bias in the orientation of the magnetic field. This could be attributed to the accretion along the 
filaments as the magnetic fields are expected to be locally preferentially tangential to these structures
\citep{bruggen05b} or to inhomogeneous radial flows in general.
The right panel of Figure 4 shows the anisotropy parameter $\beta_{v}$ for the velocity field.
The definition of this quantity is analogous to that for the magnetic fields with velocity dispersions replacing 
the magnetic fields dispersions. However, the velocity dispersion is measured with respect to the 
mean streaming velocity of the cluster. This figure bears an interesting resemblance to its magnetic counterpart. There is 
even stronger tangential bias in the velocity field in the radiative run with  anisotropic thermal conduction
than for the magnetic anisotropy parameter for the same run. 
However, there is a significant scatter in these quantities and firm conclusion about the trend could be drawn from averaging 
over many clusters. The exact values also depend on such factors as the definitions of the cluster center and its bulk velocity, and whether
the quantities are mass-weighted or not.
Nevertheless, some similarity in these quantities is not unexpected
simply due to the fact that the magnetic field is frozen to the gas and follows it. However, unlike the velocity field,
the magnetic field has a ``memory'' of past gas displacement, so the two anisotropy parameters are not expected to be identical.
Intermediate radii tend to have 
relatively
more tangential velocity field. \\
\begin{figure*}
\includegraphics[width=0.5\textwidth]{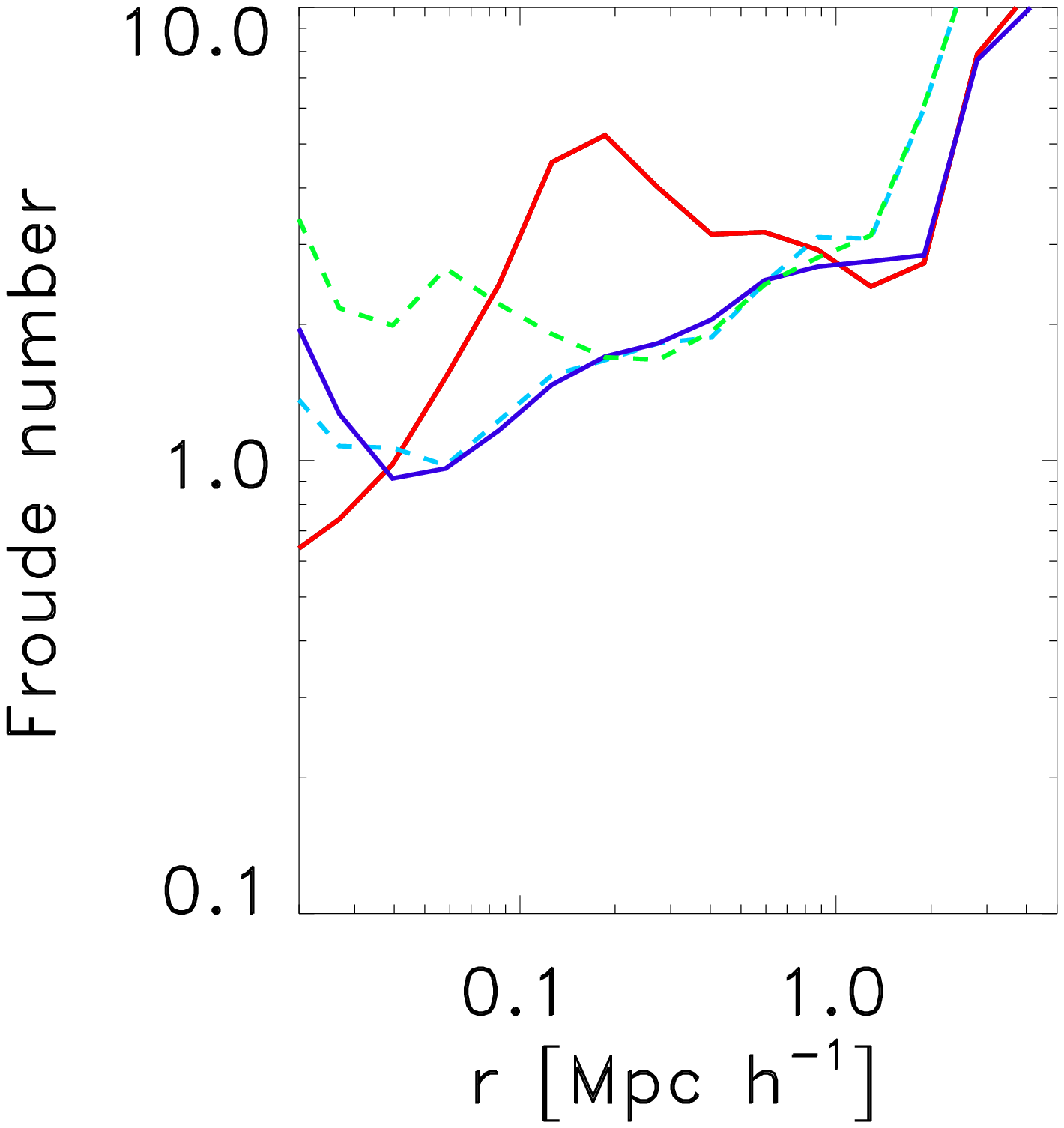}
\includegraphics[width=0.5\textwidth]{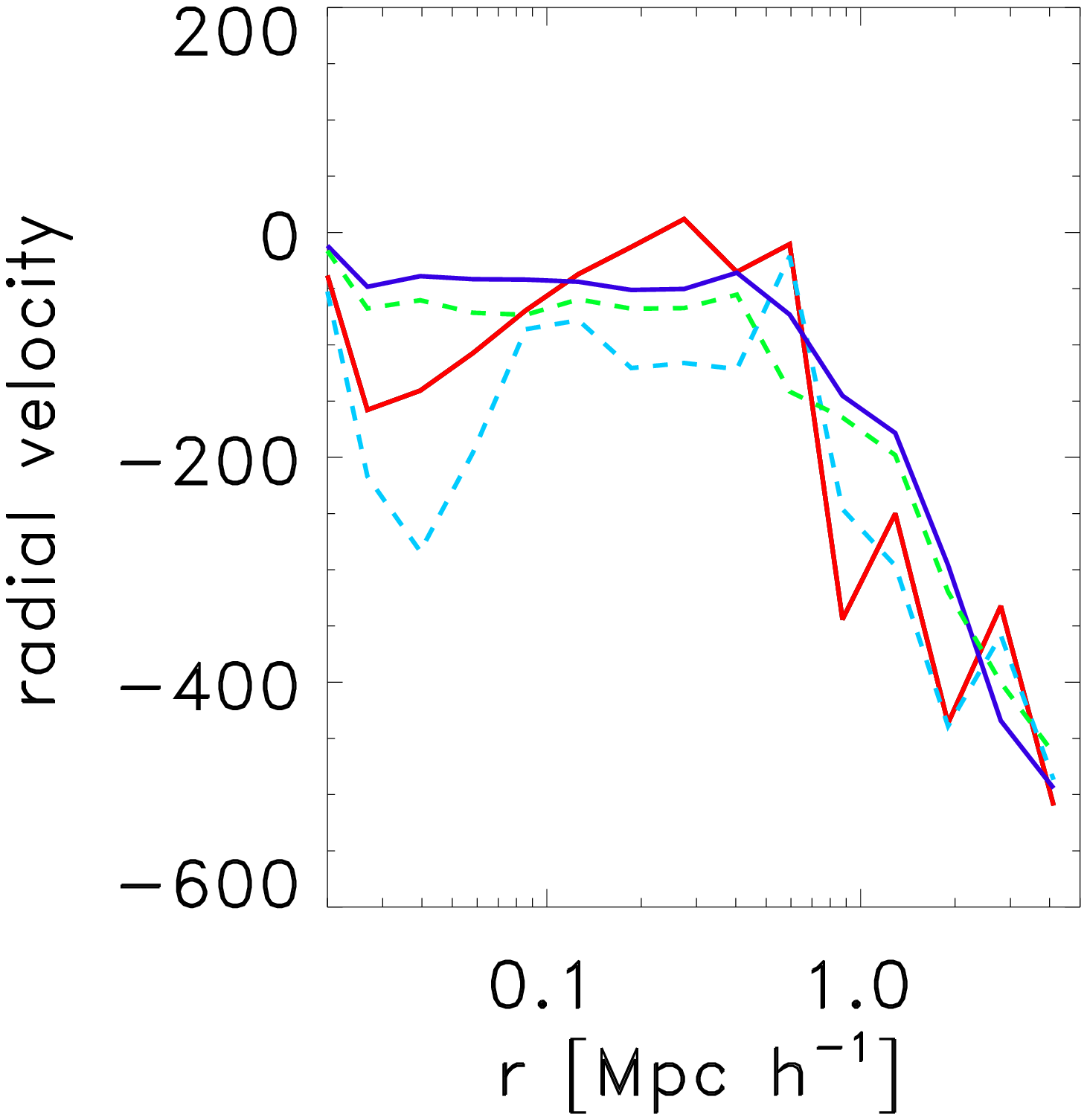} 
\caption{Froude number $Fr$ as a function of the distance from the cluster center (left panel) 
and the radial component of the gas velocity. The radial velocity plot shows net mass-weighted infall velocity. Both panels are 
plotted using 15 spherical shells of logarithmic width 0.168 in the range 15$h^{-1}$kpc $\le r\le$ 5$h^{-1}$Mpc.
The color coding of the curves is the same as in Figure 2.
For $Fr>1$ the turbulence is expected to begin to randomize the magnetic field configuration that would otherwise be established
by the instabilities.\\}
\end{figure*}
\indent
\subsubsection{Radial bias in the velocity and magnetic fields}
As argued in \citet{sharma09} and \citet{ruszkowski10}, the ability of the HBI and MTI to reconfigure the 
magnetic field depends on the level of externally driven turbulence. In the current case, the turbulence is driven by 
the structure formation motions. 
Strong turbulence can significantly affect the MTI but some residual radial bias may be present in the field.
\citet{McCourt} showed that even for strong turbulence driving, the MTI operates and significantly contributes to the velocity 
power spectrum on scales larger than the outer turbulence driving scale $L_{\rm out}$. On scales smaller than
$L_{\rm out}$ the power spectra in the conduction and adiabatic runs are nearly identical. As these isolated box experiments 
considered relatively small outer driving scale ($L_{\rm out}=h_{\rm pres}/16$, where $h_{\rm pres}$ is the pressure scaleheight
in a box $h_{\rm pres}/2$ high), larger outer driving scales may reduce the parameter space where the MTI operates.\\
\indent
If the Froude parameter is less than unity, then the turbulence does not reset 
the magnetic field configuration established by  
the MTI. We define the Froude number here as $Fr=\sigma/\omega_{BV}L$, 
where $\sigma$ is the gas velocity dispersion and $L$ is the characteristic lengthscale. Here we define it entirely for 
reference purposes as 
the pressure scaleheight $L=P/\delta P$ in  all runs. 
The characteristic frequency $\omega$ of the gas perturbed from its equilibrium state is

\begin{eqnarray}
\omega^{2} = \frac{g}{r} \left(\frac{3}{5}\left|\frac{d\ln S}{d\ln r}\right|,\left|\frac{d\ln T}{d\ln r}\right| \right),
\end{eqnarray}

\noindent
where $g$ is the total gravity and $S=k_{b}T/n_{e}^{2/3}$ is the gas entropy. The first term 
in brackets on the right side of Eq. 14 is for the standard \brunt frequency while the second term is for 
its MHD equivalent in the presence of anisotropic thermal conduction. 
The Froude number as a function of the distance from the cluster center is shown in the left panel 
in Figure 5. The meaning of the curves is the same as in Figure 2. For the runs without anisotropic conduction 
we used the standard \brunt frequency to evaluate the Froude parameter and 
for those that include anisotropic thermal conduction we used the MHD equivalent of the \brunt frequency 
(see equation 14 above). As can be seen in this figure, the Froude parameter 
typically exceeds unity. This suggests that the instabilities should be significantly affected by the turbulence.
However, the estimate of {\it Fr} is subject to uncertainties because of the arbitrariness of the choice of the 
characteristic turbulence scale. Therefore, $Fr$ parameter should only be taken as a rough guide to the impact of the turbulence.
We also experimented with other definitions of $L$ (=100 kpc, hydrostatic pressure scaleheight, and
$\sigma/|\nabla\omega|$, where $\sigma$ is the velocity dispersion and $\omega$ is the vorticity). 
In all cases typical values of $Fr$ number tend to exceed unity.\\
\indent
Despite the fact that Froude number tends to exceed unity, 
the velocity and magnetic fields indicate clear radial bias for $r\ga 0.8h^{-1}$ Mpc in Figure 4. 
At  $r\sim 0.8h^{-1}$ Mpc the result is consistent with no bias even though the timescales for the MTI development are short
(see next section). Radial bias in the magnetic field in the part of the ICM where the temperature declines with distance from the center
could be caused by the MTI. 
However, the radial bias is present at large distances from the cluster center where the contribution of turbulence to 
the pressure support is strongest, which tends to isotropize the fields.
Thus, a plausible and simple
explanation for the radial bias is  that it is due to inhomogeneous radial 
gas flows either in the bulk of the ICM or through the filaments in more distant parts of the cluster. 
We verified that indeed some residual net infall velocity is present at larger radii in all four runs, i.e., independently of whether
anisotropic thermal conduction is included. 
This effect is shown in the right panel of Figure 5 where the magnitude of the radial component of the gas velocity systematically 
increases at large distances from the center. The meaning of all curves is again the same as in Figure 2. 
This is also consistent with the radial bias in the velocity field that increases with the distance from the cluster center
(right panel in Figure 4 discussed above). 
However, there is no excess magnetic and velocity anisotropy bias in the runs with anisotropic thermal conduction.
Nevertheless, we point out that it is difficult to disentangle the effect of the MTI and the inhomogeneous radial flow
on the orientation of the magnetic fields.
Moreover, a definite statement about the likelihood of the magnetic field reorientation due to the MTI in the bulk of the cluster volume 
will have to await a systematic study of more than just one 
cluster as the dynamical states and properties of the ICM turbulence (such as the effective
outer driving scale) will be different in other clusters. 
Here we discuss the result of a pilot study that forms the basis for future investigation in this direction.\\
\indent
The radiative run without conduction (dashed light blue line, right panel in Figure 5) 
exhibits net accretion even at smaller radii which is due to the cooling flow. It is worth noting that, 
in the radiative run with anisotropic conduction, the weak net infall in the central cluster regions 
is still present but it is significantly reduced compared to the cooling-only case 
(see solid dark blue line for radiative run with conduction and dashed light blue line for the radiative run for $r<0.5h^{-1}$ Mpc).\\

\subsubsection{Observations of magnetic field orientation in clusters}
\indent
We also note that \citet{pfrommer10} reported on radio polarization measurements 
in the Virgo cluster based on the observations of magnetic field draping around the cluster galaxies.
\citet{pfrommer10} suggests that the polarization pattern in the vicinity of the magnetic ``tracer'' galaxies interacting with the ICM
is consistent with predominantly radial fields at large distances from the cluster center. They further pointed out that this 
observation is consistent with the predictions of the MTI.
For this mechanism to work, the MTI would have to either reestablish the radial magnetic field bias on a timescale at least comparable to 
the dynamical timescale of the orbiting ``tracer'' galaxies or the properties of the galaxies must be such that they do not dramatically stir
the ICM in a continuous fashion and/or that a significant time passed since the last major merger. 
Such instability growth rates may be possible but whether they are realized depends on the details of the dynamical structure of the ICM. \\

\begin{figure*}
\includegraphics[width=0.5\textwidth]{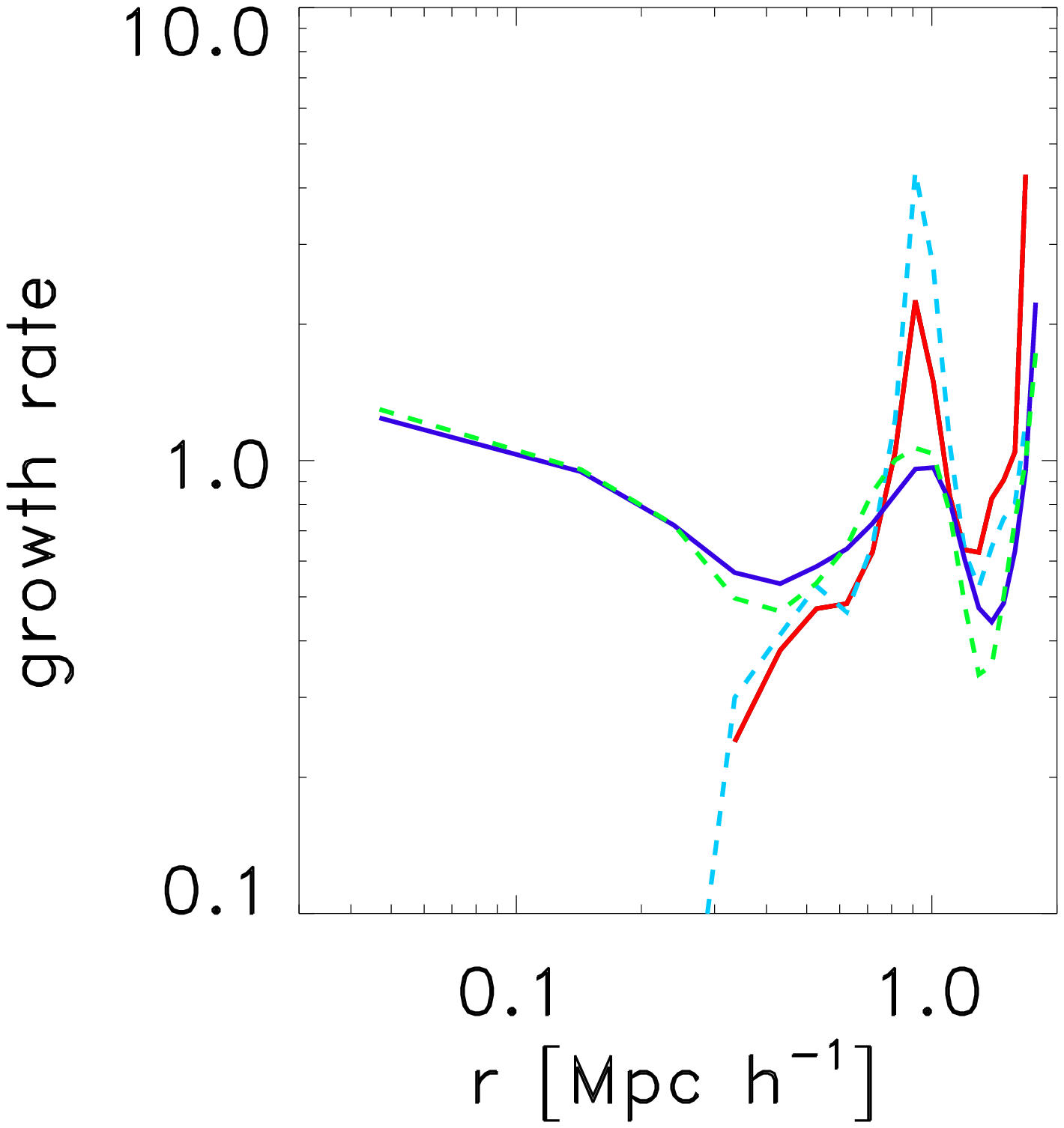}
\includegraphics[width=0.5\textwidth]{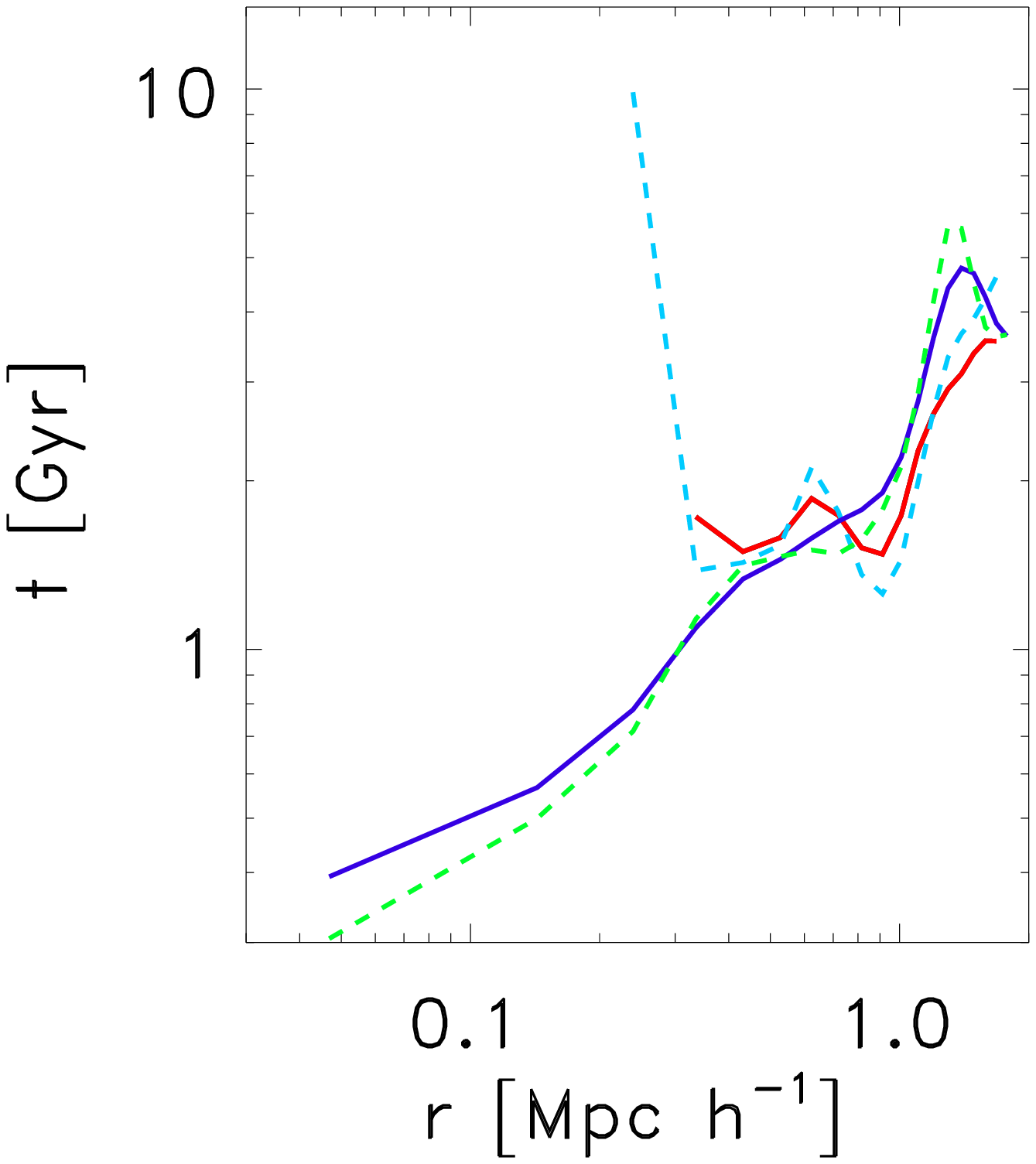} 
\caption{Left: MTI growth rates in units of \brunt frequency. Right: MTI timescales.
In both panels the color coding of the curves is the same as in Figure 2.
For the runs without anisotropic conduction the rates are inferred from the density and temperature distributions 
via the post-processing of the simulation data.\\}
\end{figure*}
\indent

\subsubsection{MTI growth rates and timescales}
In Figure 6 (left panel) we show the MTI growth rates in units of \brunt frequency.  
For the runs without anisotropic conduction, the rates shown are those that would have been present had conduction been included. 
That is, in this case the MTI rates are computed via the post processing of the simulation data.
In all cases, the growth rates were obtained from the MTI dispersion relation taking into account the slope of the temperature and entropy 
profiles. As by definition we are only interested in those fields that can be reoriented in the radial direction due to the MTI, 
we assumed tangential fields when computing the MTI growth rates. For radial fields the MTI growth rate would be infinite. 
Furthermore, we considered perturbation wavelength equal to the radius. Shorter fluctuation wavelengths would 
result in somewhat faster growth rates. The right
panel in Figure 6 shows the corresponding MTI instability growth timescales. Both panels in this figure show that 
the MTI instability has had time to develop across a wide range of distances from the cluster center in the non-radiative case.
Nevertheless, the discussion of the Froude parameter above suggests that, at least in the cluster considered here,
the instability might have been overwhelmed by the stirring motions
in the ICM. A mild radial bias in the orientation of magnetic field (and velocity) is present in the simulations at large radii but 
it is consistent with being due to either gas inflow or preferentially radial substructure motions.
However, we reiterate the point made earlier that it is difficult to disentangle the inhomogeneous radial 
flow from residual MTI.
In the runs with radiative cooling, the timescales in the more central parts of the cluster are long (dashed blue line corresponds
to high values of the MTI growth timescale for $r\la 0.2h^{-1}$ Mpc 
and the solid red line in the same region is not shown but corresponds to very long timescale).
This is caused by the 
flatness of the temperature distribution, which makes the gas neutrally buoyant. As we explain below, this 
effect leads to 
an enhanced magnetic field amplification in this region when anisotropic conduction is included in the simulation.
The same stirring motions that tend to suppress the MTI in the non-radiative case, are responsible for driving unimpeded stirring
in the radiative one, thus amplifying the magnetic fields.
However, even in the radiative runs, the MTI growth rates are reasonably short beyond $\sim 0.3h^{-1}$ Mpc from the center. We also point out
that the slopes of the temperature distribution in the runs with radiative cooling are steeper at higher redshifts, and therefore
a wider range of radii was in the unstable regime earlier on.

\subsection{Magnetic field amplification}

\begin{figure*}
\begin{center}
\includegraphics[width=0.9\textwidth]{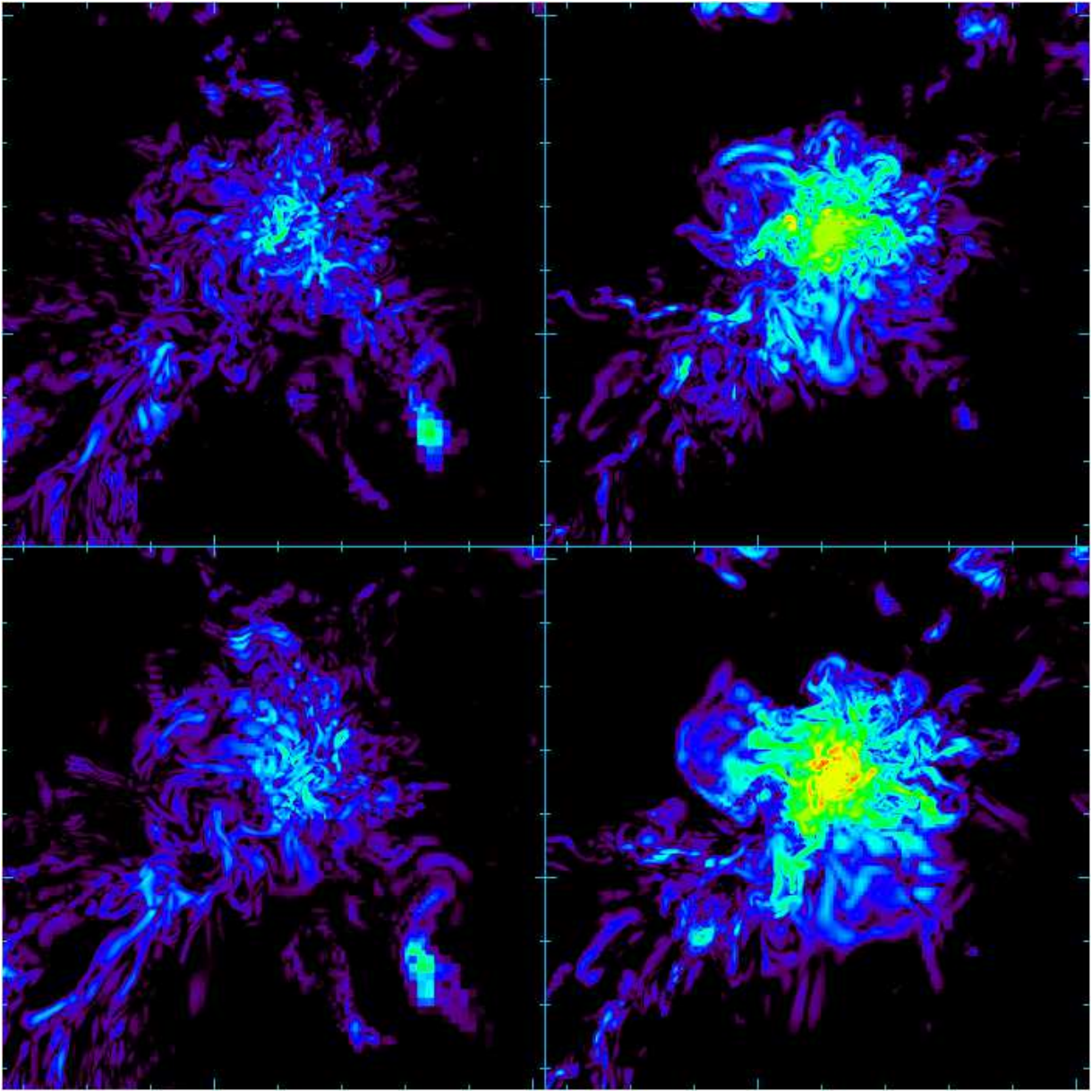}
\end{center}
\caption{Cross sections through the cluster center showing the 
distribution of the logarithm of the magnetic field pressure. The minimum and maximum range of magnetic field values
is the same in all panels. The arrangement of the figures is the same as 
in Figure 1 that shows the temperature distribution: right column is for radiative runs and bottom row is for the runs
with anisotropic thermal conduction. All panels show the central region that measures $8h^{-1}$ Mpc on the side. \\}
\end{figure*}

Cross sections through the cluster center showing the 
distribution of the logarithm of the magnetic field pressure are shown in Figure 7. The minimum and maximum range of magnetic field values
is the same in all panels. The arrangement of the figures is the same as 
in Figure 1 that shows the temperature distribution: right column is for radiative runs and bottom row is for the runs
with anisotropic thermal conduction. All panels show the central region that measures 16 Mpc on the side. 
The magnetic field is clearly amplified in the cluster center in all four cases. The amplification is much stronger in the radiative 
runs due to the compression of the cool gas and magnetic field that is frozen into it. This is consistent with the findings of 
\citet{dubois} who performed MHD simulations of cluster formation with radiative cooling. Interestingly, the radiative cooling run
with anisotropic thermal conduction (bottom right panel) shows even stronger magnetic field amplification than the radiative cooling run. 
This result is not unexpected. The additional amplification occurs in the region where the temperature gradient is significantly 
flatter than in other runs. The temperature flattening occurs outside the central cool core and up to $\sim 0.5h^{-1}$ Mpc from the cluster center. 
The nature of convective motions changes in the presence of anisotropic thermal conduction. When the temperature profile flattens, 
the ICM tends to become neutrally buoyant. This means that the restoring forces in the fluid 
diminish and the substructure infall becomes the main engine for driving the unimpeded mixing of the gas.
Consequently, this results in more efficient winding up and amplification of the magnetic field frozen into the gas via the 
kinematic dynamo effect.\\
\begin{figure*}
\includegraphics[width=0.5\textwidth]{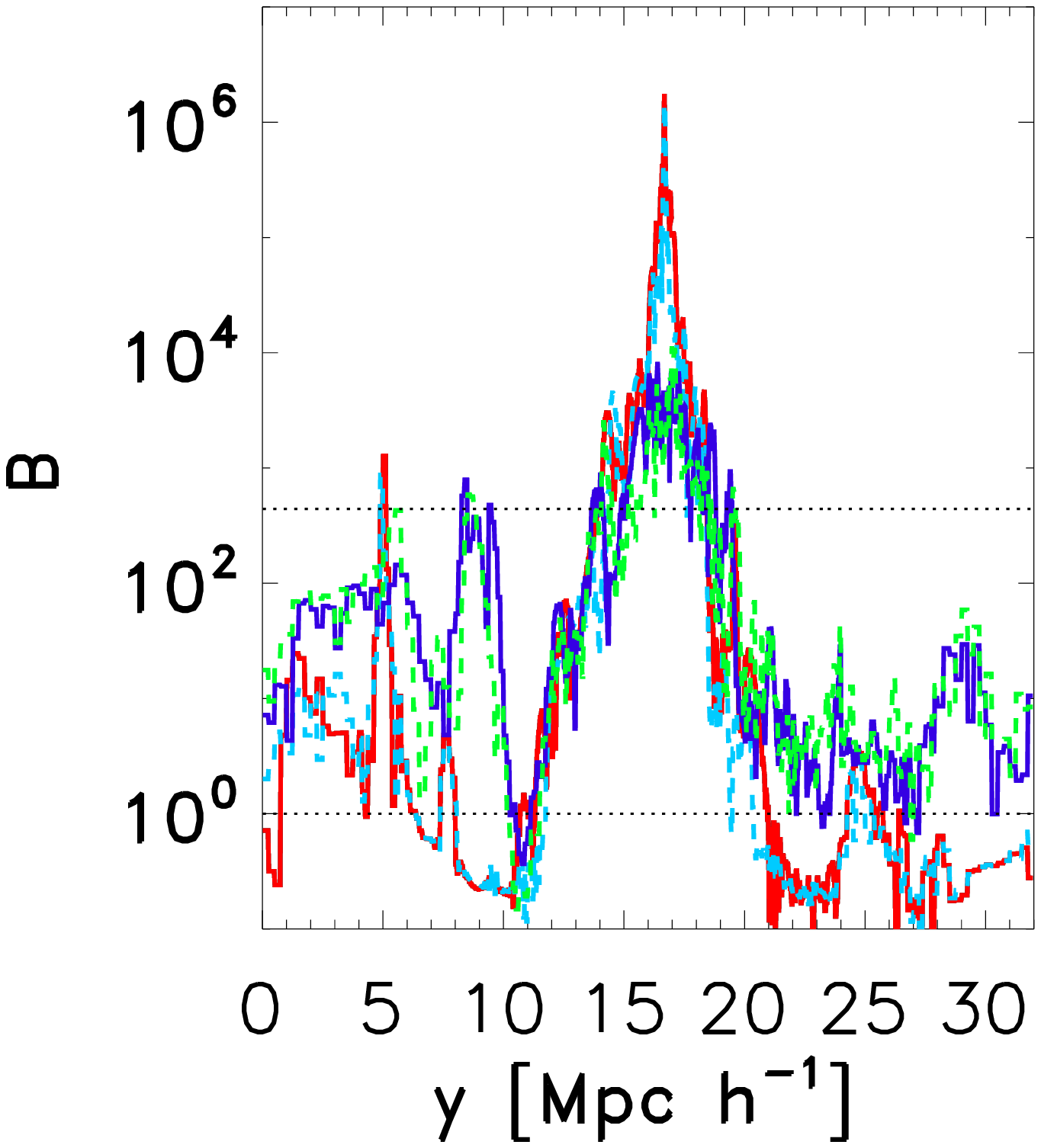}
\includegraphics[width=0.5\textwidth]{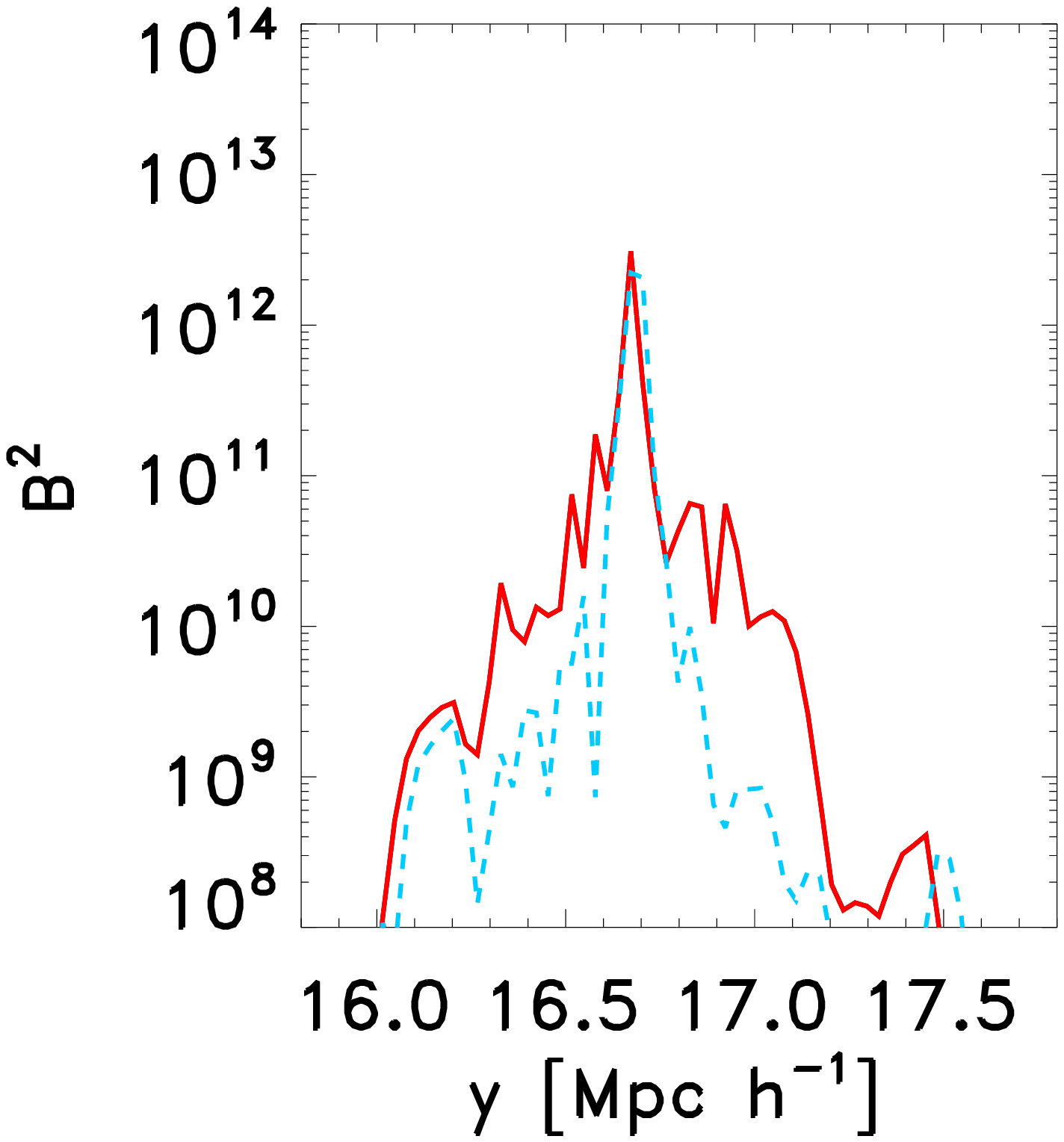}
\caption{The distribution of the magnetic field along the line passing 
through the cluster center (left panel). The color coding of the curves is the same as in Figure 2.
The top horizontal line denotes the physical field at the initial redshift ($z=20$) and the bottom one is for the value of the field that 
would result from cosmological expansion down to $z=0$ without any structure formation effects.
Right panel shows the magnetic pressure along the line passing through the cluster. Here 
the solid red line is for the anisotropic conduction and cooling while the dashed light blue line
is for the run that includes only radiative cooling.\\}
\end{figure*}
\indent
In order to better quantify the amplification of the field, we also plot the distribution of the magnetic field along the line passing 
through the cluster center. This is shown in the left panel of Figure 8. The color coding of the curves is the same as in Figure 2.
The runs with cooling boost the field by over two orders of magnitude beyond the amplification seen in the non-radiative cases.
The top horizontal line denotes the physical field at the initial redshift ($z=20$) and the bottom one is for the value of the field that 
would result from cosmological expansion down to $z=0$ without any structure formation effects. These reference levels show that 
the magnetic field in clusters is boosted beyond the initial physical field by over an order of magnitude. 
The boost in the strength of the magnetic field in the adiabatic case is $\sim 10^4$ compared to the value a uniform magnetic field
would have at $z=0$. The magnitude of this amplification is consistent with that obtained by \citet{dolag99}, \citet{dolag02} and 
in our earlier work \citep{bruggen05b} where the numerical resolution was higher. 
Simple scaling arguments show that this field strength exceeds the field expected from 
the magnetic flux freezing arguments. Specifically, for flux freezing we have for the ratio of physical field strengths, 
$B(z=0)/B(z=z_{\rm ini}) = [\rho (z=0)/\rho_{\rm ini}]^{2/3}$.
For a typical cluster, the final overdensity is $\sim 10^{3}$ above the current critical density. Thus, for the cosmology relevant to the Santa 
Barbara run we have, $B(z=0)/B(z=z_{\rm ini}) = [10^{3}/(1+z_{\rm ini})^{3}]^{2/3}\sim 0.2$, which is much smaller than the amplification factor of 
a few$\times 10$ that we see in the adiabatic simulation.
This shows that the final magnetic field in the adiabatic simulation is additionally boosted by the shearing motions during the cluster formation.
We point out that the final magnetic fields at the cluster center in the cooling run and the run with anisotropic conduction with cooling 
are boosted even further. The 
magnetic fields are $\sim 3\times 10^{3}$ times stronger than the physical field at the initial redshift of $z=20$ (or equivalently, 
over $10^{6}$ stronger than a uniform field would have at $z=0$ due to the universal expansion alone, or 
about hundred times stronger than in the adiabatic run). Such levels of magnetic field amplification 
are also in qualitative agreement with higher resolution 
simulations of \citet{dubois} who performed adiabatic and radiative runs without thermal conduction.
Such field strengths are below the MTI and HBI suppression values. Thus, for the initial field strengths considered here, the 
suppression of these instabilities by magnetic tension does not play a significant role. \\
\indent
Outside the cluster center the field at $z=0$ 
is reduced compared to its initial value. This is simply the result of the cosmological expansion. However, the field far from the
cluster center is still somewhat higher than the uniform field evolved down to $z=0$ due to the universal expansion alone.
This could be attributed to mild amplification
in filaments or smaller gravitationally bound structures. It is interesting to note that the runs with radiative cooling show
even weaker fields far from the cluster center. This is expected as the gas tends to cool down and move closer to the center of the cluster, 
filaments, and halos amplifying the field in these regions at the expense of the rest of the volume. \\
\indent
As mentioned above, the field amplification in the cooling run with anisotropic conduction shows stronger field amplification
compared to its non-conductive counterpart (Figure 7, cf. top and bottom panels in the right column). In order to better 
illustrate this effect, we zoom in on the central cluster regions and show the magnetic pressure along the line passing through the cluster 
center in the right panel of Figure 8. The solid red line is for the anisotropic conduction and cooling while the dashed light blue line
is for the run that includes only radiative cooling. It is clear from this figure that the amplification in the former run is stronger
by up to two orders of magnitude. 

\section{Conclusions}
We performed first magnetohydrodynamical simulations of cosmological galaxy cluster formation that simultaneously include 
magnetic fields, radiative cooling and anisotropic thermal conduction. The presence of anisotropic conduction 
changes the 
properties of the ICM by making it convectively unstable independently of the sign of the ICM temperature gradient. 
In our approach, we self-consistently included the amplification of the magnetic field
due to the shearing motions, gas compression enhanced by radiative cooling, and the kinematic dynamo associated with 
the anisotropic nature of conduction. Our key findings are:\\

\begin{enumerate}
\item
At large distances turbulent motions tend to reset the (radial) orientation of the 
magnetic field that the MTI tries to establish. Nevertheless, some radial bias in the orientation of the field
is seen at large radii in the runs with and without conduction. No clear excess of 
directional bias in the magnetic and velocity field is seen in in the runs with anisotropic thermal conduction.
The residual bias may be due to 
to the infall of substructures, gas accretion along the filaments and the inhomogeneous radial flows in the bulk of the ICM.
However, radial bias is also expected when the MTI operates. The degree of this bias depends on the properties of 
the turbulence driven by structure formation, and disentangling the inhomogeneous 
radial gas flows and MTI scenarios is challenging.
A systematic theoretical study of a number of clusters is required to quantify the role of MTI in shaping the 
magnetic field topology as the thermal and dynamical state of the ICM in other clusters (level of turbulent support, MTI growth rates,
effective outer scale of turbulence driven by structure formation)
will vary across the mass spectrum of clusters. Here we report on the first study of a classic {\it Santa Barbara} cluster that 
will form the basis for further investigation to address the above point.

\item
Magnetic field amplification is significantly boosted in the presence of radiative cooling which 
allows the gas to concentrate toward the cluster center. The central magnetic field at $z=0$ is amplified by over six orders of magnitude
over the value obtained without substructure formation.

\item
In the presence of anisotropic thermal conduction and radiative cooling,
the magnetic field is amplified by a kinematic dynamo process
beyond the values obtained in the cooling-only run.
This additional amplification occurs over a broad range of radii where the 
temperature profile is relatively flat. In this region, the restoring forces in the fluid 
diminish and the continuous substructure infall and stirring become 
the main engine for driving the unimpeded mixing of the gas and the amplification of the magnetic field.

\item The effective heat conduction from the hotter outer layers of the cluster to its center 
is reduced below the full Braginskii-Spitzer value. However, 
the effective radiative cooling driven accretion 
is noticeably reduced.

\item
The radiative run with anisotropic thermal conduction exhibits a tendency for a tangential bias in the velocity and magnetic fields within
the inner $\sim 0.3h^{-1}$ Mpc that could be associated with HBI or trapped gravity modes. Future higher resolution cosmological 
cluster formation studies will assess the robustness of this effect.
Any possible 
anisotropy in the magnetic field distribution may be detectable via radio polarization measurements with {\it Square Kilometer Array} and
{\it LOFAR}, while the bias in the velocity field could be probed with the future {\it International X-ray Observatory} mission.\\

\end{enumerate}

\section*{Acknowledgments}
The software used in this work was in part developed by the DOE-supported ASC/Alliance Center for
Astrophysical Thermonuclear Flashes at the University of Chicago.
MR thanks the staff NASA Ames Research center for technical help with performing the runs at the {\it Pleiades} supercomputer 
where most of the runs were performed. 
We are indebted to Chris Daley for his assistance with the particle and gravity modules in the {\it FLASH} code.
We thank Eliot Quataert, Maxim Markevitch, Christoph Pfrommer, 
Paul Nulsen, Christine Jones, Larry David, Bill Forman, Milos Milosavljevic,
John ZuHone, Mikhail Medvedev, Steve Balbus, and Fabian Heitsch for discussions.\\

\begin{figure*}
\includegraphics[width=0.5\textwidth]{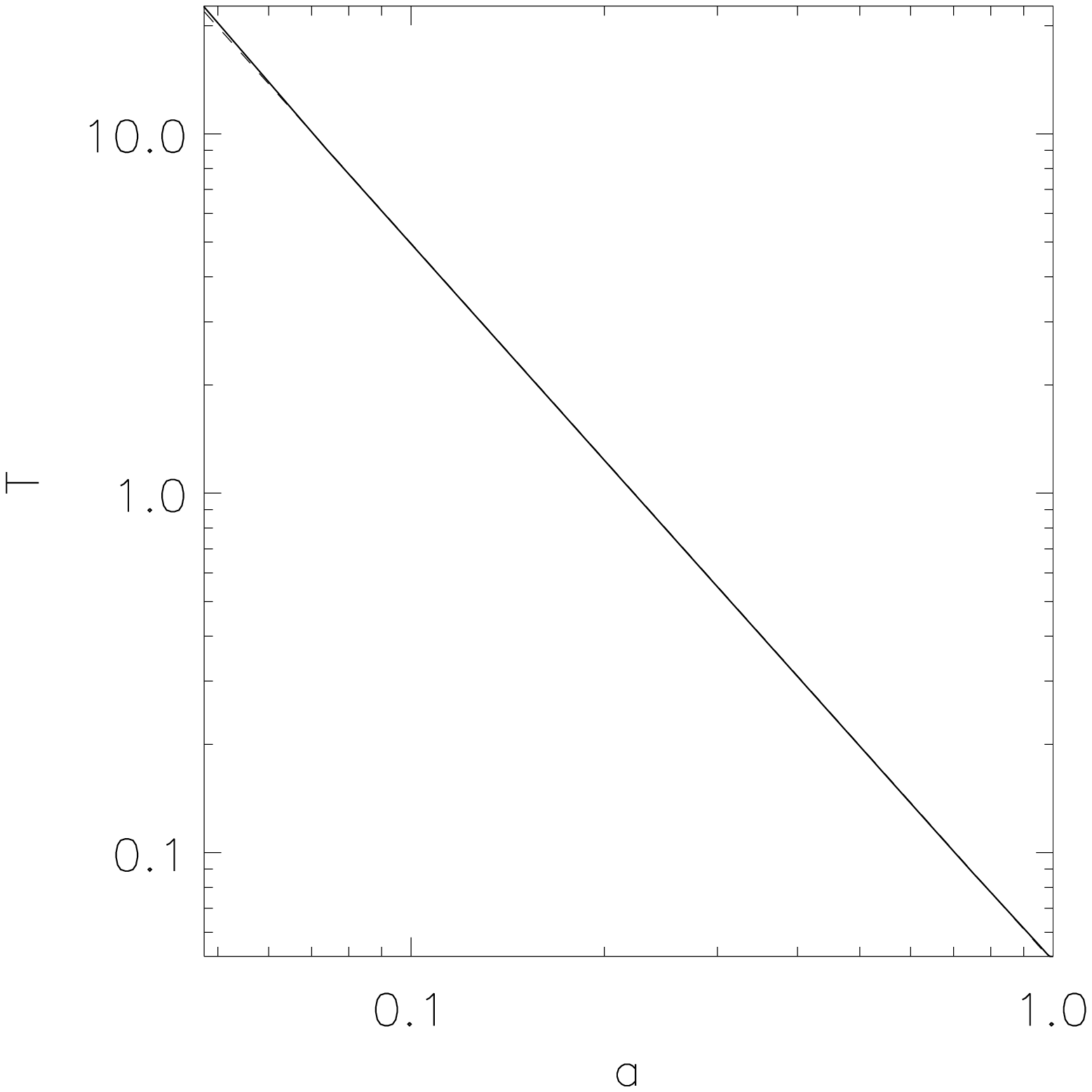}
\includegraphics[width=0.5\textwidth]{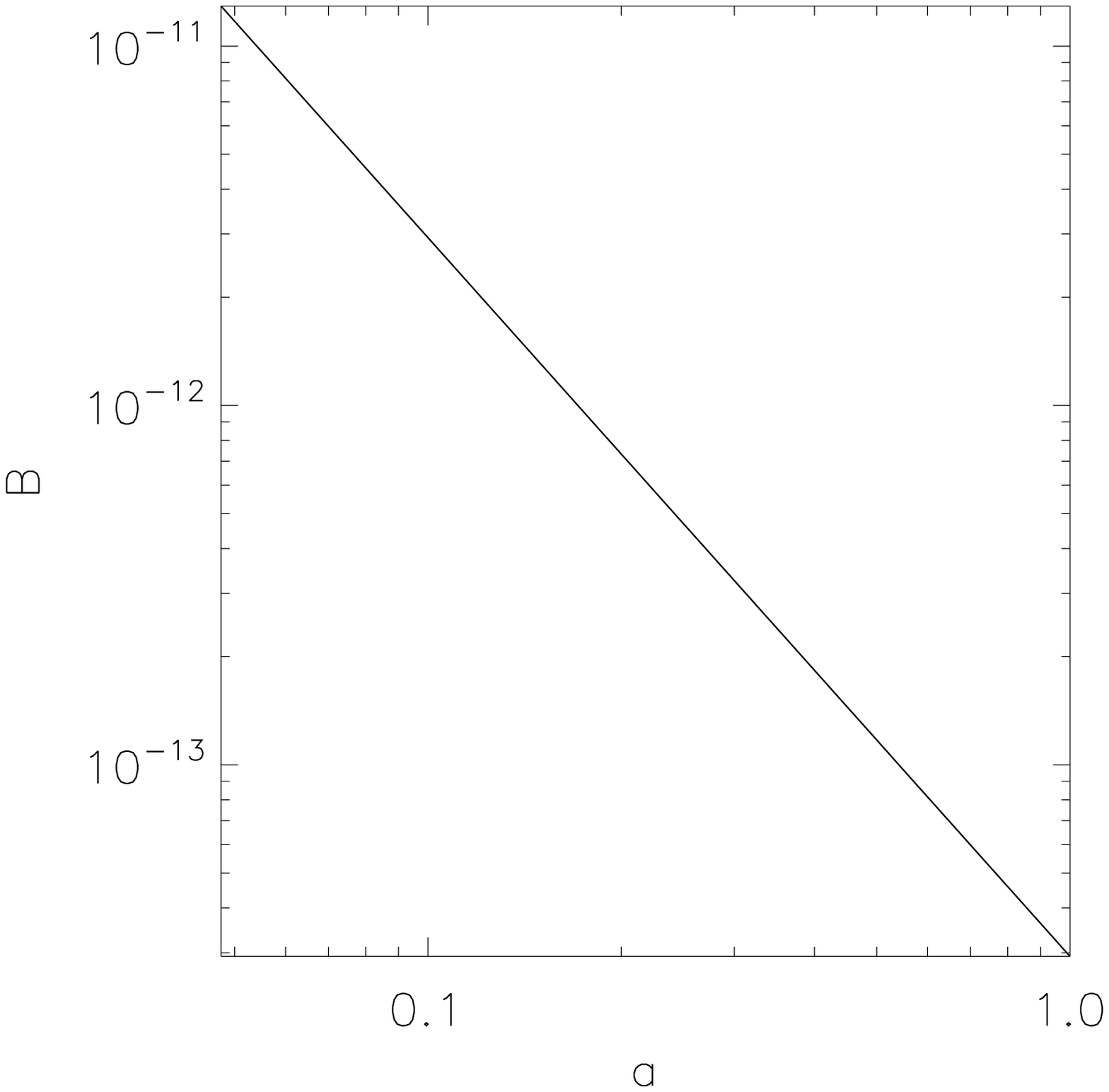}
\caption{The evolution of physical temperature (left panel) and physical magnetic field with the cosmological expansion parameter $a$.
Quantities on vertical axes in both panels are in arbitrary units. The test assumes uniform magnetic field, density and temperature
and neglects any seed fluctuations in the velocity field. The results are in perfect agreement wit the theoretical 
scaling relations (powerlaw fits to the computed relations are also plotted and match these relations perfectly.) \\}
\end{figure*}

\begin{figure}
\includegraphics*[viewport = 80 0 400 400, width = 0.5\textwidth ]{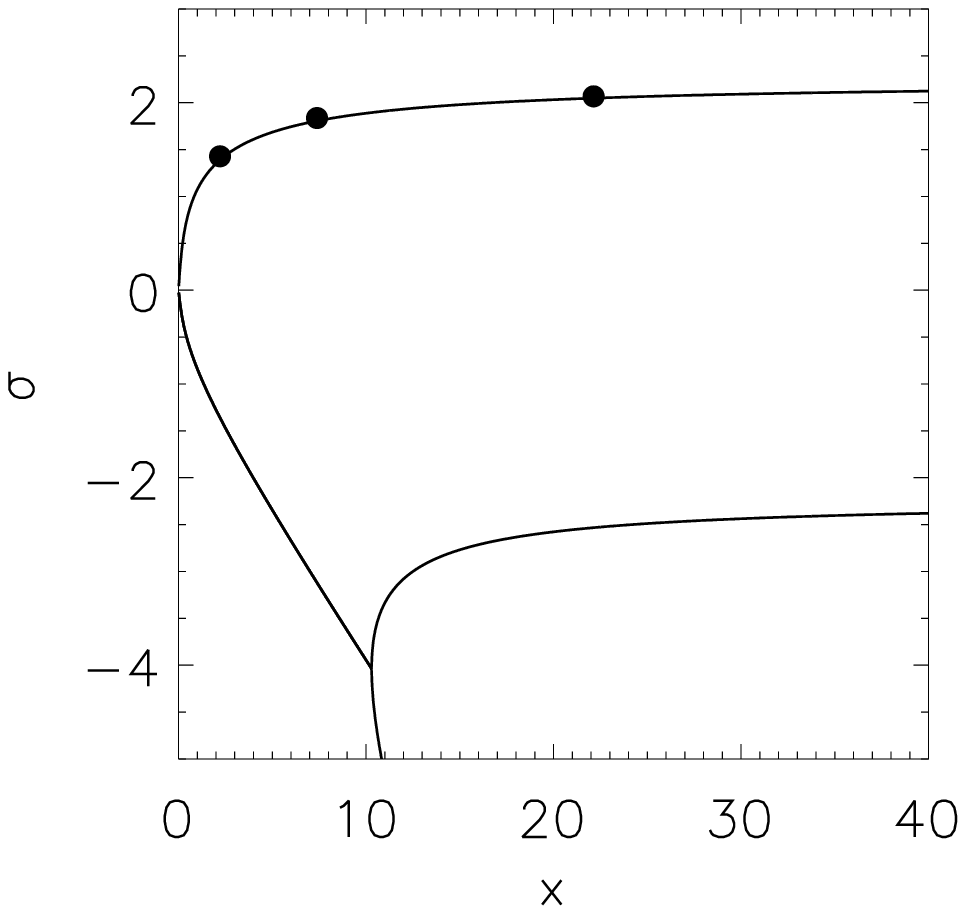}
\caption{Magnetothermal instability growth rate in units of \brunt frequency as a function of 
$x = (\gamma -1)\kappa (T/P) k^{2}/\omega_{BV}$, where $k$ is the wavenumber of the velocity fluctuation and 
$\omega_{BV}$ is the \brunt frequency.
Solid line denotes the prediction from linear theory and the dots show the results obtained with the 
{\it FLASH} code.\\}
\end{figure}

\section{Appendix A}
In order to check the implementation of the cosmological terms in the MHD equations we evolved
spatially constant matter density and magnetic field while neglecting any velocity perturbations. The result of this test is shown 
in Figure 9. In the left panel we show the scaling of the temperature (in arbitrary units) with the cosmological expansion factor $a$.
Shown are the code result for the physical temperature 
and the power law fit (solid and dashed lines are practically indistinguishable). The slope of $T(a)\propto a^{-2}$
agrees with the standard theoretical expectation. The right panel shows the physical magnetic field in arbitrary units as a function of the
scaling parameter $a$ and a powerlaw fit to this relation, both as a function of the scaling parameter $a$. Here again the fit is perfect, 
the solid and dashed lines overlap, and the slope 
of the field is ($B(a)\propto a^{-2}$) ensures the conservation of the magnetic flux.

\section{Appendix B}
In order to test the implementation of the anisotropic thermal conduction module we compared 
linear theory MTI growth rates with the code results. This test is very similar to the one discussed in 
Parrish \& Stone (2008). That is, we set up a two dimensional stratified hydrostatic atmosphere with very shallow
density and temperature profiles such that

\begin{eqnarray}
T(z)=T_{o}(1-y/y_{o})
\end{eqnarray}
\begin{eqnarray}
\rho(z)=\rho_{o}(1-y/y_{o})^{2},\\ \nonumber
\end{eqnarray}

\noindent
where $T_{o}$ and $\rho_{o}$ are constants. The characteristic lengthscale $y_{o}$ was set to 1\% of the horizontal 
height of the computational box. The gravitational field was assumed constant throughout the computational domain.
We set hydrostatic boundary conditions on the top and bottom boundary. We also impose constant temperature in the boundary zones to 
prevent the escape of thermal energy from the simulation box via thermal conduction. That is, the computational domain is effectively
adiabatic. We set periodic boundary conditions in the horizontal direction. Initial magnetic field is horizontal with very high magnetic 
$\beta$ parameter. In order to seed the instability we introduce a very small sinusoidal velocity perturbation in the vertical direction.
Specifically, the initial vertical component of the velocity field is given by

\begin{eqnarray}
v_{z}(y)=v_{o}\sin(2\pi y/L),\\ \nonumber
\end{eqnarray}

\noindent
where $v_{o}$ is a very small constant amplitude of the velocity perturbation and $L$ is the box size in the horizontal $y$-direction.
We evolve such initial conditions and, using the velocity field as a function of time, we 
compute the instability growth rate $\sigma$ following a prescription similar to that in \citet{parrish05}.
These results are then compared to the linear theory prediction in the nondimentional form

\begin{eqnarray}
\sigma^{3}+\frac{1}{\gamma}\sigma^{2}x+\sigma + \frac{d\ln T}{d\ln S}x = 0,\\ \nonumber
\end{eqnarray}

\noindent
where $\sigma$ is the instability growth rate in units of the \brunt frequency $\omega_{BV}$ and

\begin{eqnarray}
x = (\gamma -1)\kappa\frac{T}{P}k^{2}\omega_{BV}^{-1},\\ \nonumber
\end{eqnarray}

\noindent
where $k$ is the wavenumber of the velocity fluctuation.
The result of this test is shown in Figure 10. The growth rates predicted by the code are in excellent agreement with the theoretical
expectations.\\

\newpage
\bibliography{master_references} 

\label{lastpage}

\end{document}